\documentclass[12pt,a4paper]{article}
\pdfoutput=1

\usepackage{amssymb}
\usepackage{amsmath}
\usepackage{amsfonts}
\usepackage{authblk}
\usepackage{setspace}
\usepackage{color}
\usepackage[pdftex]{graphicx} % For jpg etc figures under pdflatex.
\usepackage{placeins}

\numberwithin{equation}{section}

\newcommand{\BbbR}{\mathbb{R}}

\DeclareMathOperator{\Realpart}{Re}

\DeclareMathOperator{\sgn}{sgn}

\DeclareMathOperator{\arccot}{arccot}

\DeclareMathOperator{\Ci}{Ci}

\title{Smooth and sharp creation of a Dirichlet wall\\ 
in 1+1 quantum field theory: 
how singular\\ 
is the sharp creation limit?}

\author{Eric G. Brown}
\affil{Department of Physics and Astronomy, University of Waterloo, 
Waterloo, Ontario N2L 3G1, Canada}
\author{Jorma Louko}
\affil{School of Mathematical Sciences, 
University of Nottingham,\\ 
Nottingham NG7 2RD, 
UK}

\date{{\small April 2015, revised July 2015}\\[1ex]
{\small Published in JHEP {\bf 1508} (2015) 061\footnote{Post-publication note:
In Section~\ref{sec:Minkowski}, 
$\langle T_{uu}\rangle$ \eqref{eq:Tuu-Minkowski}
tends to $\infty$ as $u \to \lambda^{-1}_-$, so fast that 
$\langle E_{\text{tot}} \rangle$ in 
\eqref{eq:Ttot-Mink-def} and 
\eqref{eq:Etot-Mink-scaled}
equals~$\infty$, under mild technical
assumptions about  
$h(y)$~\eqref{eq:theta-scaled}. 
Equation \eqref{eq:Etot-Mink-as} is hence incorrect in that the 
term denoted therein by $O(1)$ equals $\infty$. 
For related discussion, 
see arXiv:1610.08455v2. 
Similar comments may apply to 
\eqref{eq:DeltaTuu-cavity}, 
\eqref{eq:Etot-cavity-scaled}
and \eqref{eq:Etot-cavity-as}
in Section~\ref{sec:cavity}. 
The results about detector response versus total energy are unaffected since 
they are obtained with the boundary condition family 
\eqref{eq:h-incomplete} rather than~\eqref{eq:theta-scaled}.}}}

\begin{document}

\maketitle
\begin{abstract}
We present and utilize a simple formalism for the smooth 
creation of boundary conditions within relativistic quantum 
field theory. We consider a massless scalar field in 
$(1+1)$-dimensional flat spacetime 
and imagine smoothly transitioning from there being 
no boundary condition to there being a two-sided Dirichlet mirror. 
The act of doing this, expectantly, generates a flux of real 
quanta that emanates from the mirror as it is being created. 
We show that the local stress-energy tensor of the flux 
is finite only if an infrared cutoff is introduced, 
no matter how slowly the mirror is created, 
in agreement with the perturbative results of Obadia and Parentani. 
In the limit of instantaneous 
mirror creation the total energy injected into the field becomes
ultraviolet divergent, but the response of an Unruh-DeWitt 
particle detector passing through the infinite burst of 
energy nevertheless remains finite. 
Implications for vacuum entanglement extraction and 
for black hole firewalls are discussed. 
\end{abstract}

% \singlespacing

\section{Introduction\label{sec:intro}}

Within the realm of relativistic quantum field theory, both in flat and curved spacetimes, the study of time-dependent boundary conditions has been a staple exercise in understanding particle-creation phenomena~\cite{birrell-davies}. A non-inertially moving mirror, for example, induces the production of real particles out of the vacuum. Within a cavity setting this is commonly referred to as the dynamical Casimir effect~\cite{Moore70}, in which rapidly varying the length of an optical cavity can dynamically generate photons. This effect has been experimentally verified with a cQED analogue 
system~\cite{Wilson11}. Recently, there has been an increasing interest in utilizing the effect for quantum information processing and quantum metrology~\cite{Ahmadi14,Benenti14,Brown14}.

The majority of the existing literature is focused on the effects of moving boundaries. Here, we wish to properly examine a somewhat different case, and one that has been gaining interest in a number of areas. Rather than moving a boundary, we will instead create one. In particular, we take a $1+1$ dimensional massless scalar field and consider at the origin a self-adjointness boundary condition that transitions smoothly in time between there being no boundary to there being a two-sided Dirichlet wall. Physically, one can imagine such a procedure being implemented via a reflectivity-tunable barrier~\cite{Hastings05}. Unsurprisingly, such a procedure also generates quanta out of the vacuum that radiate away from the creation event. Our goal in this paper is to examine the stress-energy contribution to the field and the response of a particle detector.
As part of this exposition we will take the limit of instantaneous wall creation.

There are several motivations behind studying such a scenario. For
example, as has been pointed out by Unruh~\cite{unruh-seoultalk}, the
act of instantaneously creating a mirror produces
regions of spacetime between which field correlations cannot
propagate.
On the horizon
separating these regions (the future lightcone of the creation event)
there is expected to be a flux of quanta of diverging energy density
and diverging total energy (as we will confirm). Interestingly, this
phenomenon is very analogous to the much-debated black hole firewall
\cite{Almheiri:2012rt,Susskind:2013tg,Almheiri:2013hfa,Page:2013mqa,Almheiri:2013wka,Hotta:2013clt,Harlow:2014yka}
and related constructs
\cite{braunstein-et-al,Mathur:2009hf,Hutchinson:2013kka} in which lack
of correlation between the inside and outside of a black hole is
proposed to induce a violent horizon. Indeed, artificially
constructing uncorrelated spatial regions has been used as a
simplified firewall
model~\cite{Louko:2014aba,Martin-Martinez:2015dja}. By considering the
instantaneous limit of mirror creation within our formalism we are
able to gain further insight into the nature of the divergence
associated with firewalls.

The rapid creation of a mirror has recently gained further interest in 
studying the nature of vacuum entanglement~\cite{Brown:2014qna,Asplund:2013zba,Asplund:2014coa}. 
It was shown in \cite{Brown:2014qna}
that the two bursts of quanta produced by introducing a mirror are 
entangled with each other, and that this entanglement derives exactly 
from the previously present vacuum entanglement. 
The UV-divergent energy of these bursts is seen to be equivalent to the 
UV-divergence of the entanglement entropy between connected regions. 
This protocol has been proposed as a means of experimentally 
verifying vacuum entanglement. 
In any real experiment, however, the introduction of the mirror 
will take place over a finite time interval. 
In addition to theoretical insights into the sharp limit, 
considering a smooth transition (as we do here) 
may therefore prove vitally important for the development of such a program.

We have several goals in the current work, and give several different
results of interest. 
First, we present a formalism for analysing 
a quantised massless scalar field
in $(1+1)$-dimensional flat spacetime 
under time-dependent boundary conditions that are at each instant of time 
given by a specific self-adjoint extension 
of the spatial part of the wave equation \cite{reed-simonII,blabk,Bonneau:1999zq}, 
building on
previous treatments in a variety of contexts
\cite{Anderson:1986ww,trousers-revisited,Jaekel:1993np,Balachandran:1995jm,Marolf:1996gn,Obadia:2001hj,Johansson2010,Silva:2011fq,Farina:2012qd,Rego:2013bba,Rego:2014wta,Doukas:2014bja}. 
We use this formalism to analyse the smooth creation of a Dirichlet
wall, both in full Minkowski space and at the centre of a Dirichlet cavity. We in
particular compute the renormalized stress-energy expectation value 
in the quantum state in which no particles are present before the wall starts to form. 
In full Minkowski space, we find that the stress-energy 
is infrared divergent everywhere on the light cone of the evolving wall, 
no matter how slow the change in the boundary condition, 
as was previously observed within the perturbative 
analysis of \cite{Obadia:2001hj} 
(for related observations in the back-reaction context see~\cite{Jaekel:1993np}): 
in full Minkowski space it is hence necessary to 
introduce an infrared cutoff by hand. 
For a wall that is forming within a cavity, by contrast, 
the stress-energy is finite without additional cut-offs since the cavity already provides 
an effective infrared cutoff. 

Second, we consider the limit of instantaneous wall creation, 
by taking to zero the time interval over which the wall is created, 
while keeping fixed
the dimensionless profile function by which the boundary condition evolves within this interval. 
We show that in this limit the stress-energy tensor vanishes everywhere except 
on the light cone of the wall creation event, 
but the limit is too singular for the stress-energy to be describable as 
a well-defined distribution with support on the light cone of the wall creation event, and in particular the total energy emitted into the field diverges. 
These outcomes are consistent with
the instantaneous wall creation
discussion in~\cite{unruh-seoultalk}, with the instantaneous topology
change discussion in~\cite{Anderson:1986ww,trousers-revisited},
with the perturbative discussion in \cite{Obadia:2001hj}
and with the conformal field theory discussion
in~\cite{Asplund:2013zba}. 

Third, we 
compute the response of an Unruh-DeWitt particle detector
\cite{unruh,dewitt} that couples linearly to the proper time
derivative of the field
\cite{Louko:2014aba,Raine:1991kc,Raval:1995mb,Davies:2002bg,Wang:2013lex,Martin-Martinez:2014qda,Juarez-Aubry:2014jba,Hotta:2015yla},
choosing the derivative coupling because it is less sensitive to the
infrared ambiguity of the Wightman function of a $(1+1)$-dimensional
massless field~\cite{Juarez-Aubry:2014jba}. 
We take the detector to move inertially in full Minkowski spacetime. 
Working within first
order perturbation theory, we find that in the instantaneous wall
creation limit the detector's response has two surprising properties.
First, the response remains \emph{finite\/}, despite the divergent
total energy through which the detector passes.  Second, the response
depends on the infrared cutoff, even though the response in a number
of other states, including the Minkowski vacuum, is independent of the
infrared ambiguity~\cite{Juarez-Aubry:2014jba}. 
These results are similar to what was found in 
\cite{Louko:2014aba,Martin-Martinez:2015dja} for detectors coupled to a Minkowski 
spacetime model of a black hole firewall~\cite{Almheiri:2012rt}, 
and they add to the evidence that 
material systems modelled by the Unruh-DeWitt detector are significantly less sensitive 
to quantum field theoretic singularities than might be 
anticipated by considering just the local stress-energy of the field. 

This paper is organized as follows. We begin in Section
\ref{sec:Minkowski} with an introduction to the formalism and fully
work out the evolution of the quantum field for wall creation that
takes place smoothly over a finite interval of time in Minkowski
space. We compute the stress-energy associated with this process, 
inserting by hand an infrared cutoff, 
and we show that the total energy diverges in the sharp creation limit. In
Section \ref{sec:cavity} we perform the same analysis in the case of a
Dirichlet cavity, 
demonstrating that the cavity acts as an infrared cutoff. 
In Section \ref{sec:example}
we show that similar properties hold for creating a wall in Minkowski
space over an infinite interval of time with a specific profile that
allows computations to be done in terms of elementary functions. In
Section \ref{sec:detector-on-Minkowski} we go on to use this specific
profile to analyse an inertial particle detector and to demonstrate,
among other results, the response to remain finite even in the
sharp-creation limit. 
Technical material is deferred
to Appendices \ref{app:sa-linewithpoint}--\ref{app:lambdalimit}. 
Appendix \ref{app:bogocoeffs} presents a brief discussion of the
wall creation in terms of Bogoliubov coefficients, 
both in Minkowski space and in the cavity. 

We denote complex conjugation by an overline.  $O(x)$ denotes a
quantity such that $O(x)/x$ remains bounded as $x\to0$,
$O^{\infty}(x)$ denotes a quantity that goes to zero faster than any
positive power of $x$ as $x\to0$, and $O(1)$ denotes a quantity that
remains bounded in the limit under consideration.

\section{Wall creation 
in Minkowski spacetime\label{sec:Minkowski}} 

\subsection{Classical field}

We work in $(1+1)$-dimensional Minkowski spacetime, 
with standard global Minkowski coordinates $(t,x)$, 
in which the metric reads $ds^2 = - dt^2 + dx^2$. In the global null coordinates 
$u = t-x$ and $v=t+x$, 
the metric reads $ds^2 = - du \, dv$. 

We consider a real massless scalar field~$\phi$. 
Without a wall, the field equation is the
Klein-Gordon equation, 
\begin{align}
\partial_t^2\phi  - \partial_x^2 \phi =0
\ , 
\label{eq:KG-standard}
\end{align}
where $- \partial_x^2$ has its usual meaning as
an essentially self-adjoint positive definite operator on $L_2(\BbbR)$. 

To introduce a wall at $x=0$, 
we replace \eqref{eq:KG-standard} with 
\begin{align}
\partial_t^2\phi  - \Delta_{\theta(t)} \phi =0
\ , 
\label{eq:KG-wallgen}
\end{align}
where $\{- \Delta_\theta \mid \theta \in [0, \pi/2] \}$ 
is the one-parameter family of self-adjoint extensions of 
$- \partial_x^2$ on $L_2(\BbbR\setminus\{0\})$ described in 
Appendix~\ref{app:sa-linewithpoint}. 
As indicated in~\eqref{eq:KG-wallgen}, we allow 
$\theta$ to depend on~$t$. 

The special case $- \Delta_{\pi/2}$ is that of the unique 
self-adjoint extension of 
$- \partial_x^2$ on $L_2(\BbbR)$, corresponding to no wall at $x=0$. 
The special case $- \Delta_0$ is that of an 
impermeable wall at $x=0$ with the Dirichlet boundary condition on each side. 
For the intermediate values of~$\theta$, 
$- \Delta_\theta$ interpolates
between these two extremes, 
involving no boundary conditions for spatially odd wave functions but a 
two-sided Robin boundary condition 
[equation \eqref{eq:app:robin} in Appendix~\ref{app:sa-linewithpoint}]
for spatially even wave functions. 

The spectrum of each $- \Delta_\theta$ is the positive continuum. 
The wave equation \eqref{eq:KG-wallgen}
is hence free of tachyonic instabilities 
and provides a viable starting point for quantisation. 

In physics terms, 
the wave equation \eqref{eq:KG-wallgen} can be written for $0 < \theta \le \pi/2$ 
as 
\begin{align}
\left[
\partial_t^2  - \partial_x^2 + \frac{2 \cot\bigl(\theta(t)\bigr)}{L} \, \delta(x)
\right]
\phi =0
\ , 
\label{eq:KG-wall-pot}
\end{align}
where $\delta(x)$ is Dirac's delta-function and 
the positive constant $L$ of dimension length 
is as introduced in Appendix~\ref{app:sa-linewithpoint}. 
The wall at $x=0$ corresponds hence to a potential term proportional 
to $\delta(x)$ with a $\theta$-dependent coefficient. 
The coefficient is positive for $0<\theta<\pi/2$, 
and it tends to $0$ in the no-wall limit $\theta\to(\pi/2)_-$ 
and to $+\infty$ 
in the Dirichlet wall limit $\theta\to0_+$.

In the rest of this section we assume that $\theta(t)$ 
interpolates between no wall and a fully-developed 
Dirichlet wall over a finite interval of time.
We may assume without loss of generality that the wall 
creation begins at $t=0$, and we write the moment at which the 
Dirichlet wall is fully formed as $t = \lambda^{-1}$ where $\lambda>0$. 
We parametrise $\theta(t)$ as 
\begin{align}
\theta(t) = 
\begin{cases}
\pi/2 & \text{for $t \le 0$}
\ , 
\\
\arccot \! \left[\lambda L \cot\bigl(h(\lambda t)\bigr)\right] 
& \text{for $0 < t < \lambda^{-1}$}
\ , 
\\
0 & \text{for $t \ge  \lambda^{-1}$}
\ , 
\end{cases}
\label{eq:theta-param}
\end{align}
where
$h: \BbbR \to \BbbR$ 
is a smooth function such that 
\begin{subequations}
\label{eq:theta-scaled}
\begin{alignat}{2}
& h(y) = \pi/2 &\quad& \text{for $y\le0$}
\ ,  
\\
& 0 < h(y) < \pi/2 && \text{for $0 < y < 1$}
\ , 
\\
& h(y) = 0 && \text{for $y\ge1$}
\ . 
\end{alignat}
\end{subequations}
Over the interval $0<t< \lambda^{-1}$, the 
boundary condition \eqref{eq:app:robin} 
then reads 
\begin{align}
\lim_{x\to0_{\pm}}
\frac{\partial_x \phi(t,x)}{\phi(t,x)} 
= \pm 
\lambda \cot\bigl(h(\lambda t)\bigr)
\ . 
\label{eq:scaled-BC}
\end{align}
The parametrisation $\eqref{eq:theta-param}$ hence means 
that $\lambda^{-1}$ is the length of the time interval 
over which the boundary condition \eqref{eq:scaled-BC} 
evolves into Dirichlet, while the dimensionless function 
$h$ specifies the shape of the evolution in \eqref{eq:scaled-BC}  
over this time interval. 
The limit in which a wall is created rapidly but 
the shape of the evolution is held fixed 
is the limit of large $\lambda$ with fixed~$h$. 

We emphasise that the coefficient of $\delta(x)$ 
in \eqref{eq:KG-wall-pot} tends to $+\infty$ when the wall 
becomes a fully-developed Dirichlet wall, 
but the above description in terms of $\theta(t)$ 
nevertheless provides a control of the smoothness 
of this approach to the Dirichlet wall, 
and we shall verify in subsection \ref{subsec:Mink-modefunctions} 
below that the mode functions indeed remain smooth even when 
the Dirichlet wall is reached. 
It would be possible to consider alternatives to~\eqref{eq:KG-wall-pot}, 
such as \cite{Obadia:2001hj}
\begin{align}
\left[
\partial_t 
\bigl(1 - \Lambda(t) \delta(x) \bigr) 
\partial_t 
- \partial_x^2 
\right]
\phi =0
\ , 
\label{eq:KG-wall-timedd}
\end{align}
where $\Lambda(t) =0$ for $t\le0$ and $\Lambda(t) \to +\infty$ 
as $t \to \lambda^{-1}_-$; in particular, a potential advantage of 
\eqref{eq:KG-wall-timedd} is that the wall is softer in the 
infrared, with implications for the stress-energy tensor~\cite{Obadia:2001hj}. 
For \eqref{eq:KG-wall-timedd}, one would however need to investigate 
anew the conditions on $\Lambda(t)$ 
as $t \to \lambda^{-1}_-$ to guarantee an appropriate sense of 
smoothness on reaching the Dirichlet wall.

\subsection{Mode functions\label{subsec:Mink-modefunctions}}

As preparation for quantisation, 
we need to find the mode solutions that reduce to the usual 
Minkowski modes for $t\le0$, where the wall has not yet started to form. 

Since the spatially odd solutions to the field equation 
\eqref{eq:KG-wallgen} do not feel the wall, 
it suffices to consider the spatially even solutions. 
It further suffices to write down the expressions 
for these solutions in the half-space $x>0$; 
by spatial evenness, the expressions at 
$x<0$ follow by 
$(t,x) \mapsto (t,-x)$, or in terms of the null coordinates, 
by $(u,v) \mapsto (v,u)$. 

We work in the null coordinates $(u,v)$ and look for the mode solutions
with the ansatz 
\begin{align}
U_k(u,v) = \frac{1}{\sqrt{8\pi k}} 
\left[
e^{-ikv} + E_k(u)
\right]
\ , 
\label{eq:Umode-ansatz}
\end{align}
where $k>0$ and $E_k$ is to be found. 
Each term in \eqref{eq:Umode-ansatz} satisfies the wave equation at $x>0$, 
and the left-moving part of 
$U_k$ has the standard form proportional to~$e^{-ikv}$. 

Requiring 
\eqref{eq:Umode-ansatz} to satisfy \eqref{eq:app:robin-plusside}
with $\theta=\theta(t)$ gives 
\begin{align}
L \sin\bigl(\theta(t)\bigr) 
\frac{d}{dt}
\! \left[
e^{-ikt} - E_k(t)
\right]
&= \cos\bigl(\theta(t)\bigr) 
\! \left[
e^{-ikt} + E_k(t)
\right]
\ . 
\label{eq:E-ev-eq}
\end{align}
With $\theta(t)$ parametrised by~\eqref{eq:theta-param}, the solution is 
\begin{align}
E_k(u) = R_{k/\lambda}(\lambda u)
\ , 
\label{eq:E-intermsof-R}
\end{align}
with 
\begin{align}
R_K(y) = 
\begin{cases}
{\displaystyle{e^{-iK y}}}
& \text{for $y \le 0$}
\ , 
\\
{\displaystyle{e^{-iK y} - \frac{2}{B(y)}
\int_0^{y} {B'}(y') \, e^{-iK y'} \, dy'}}
& \text{for $0 < y < 1$}
\ , 
\\[1ex]
{\displaystyle{-e^{-iK y}}}
& \text{for $y \ge 1$}
\ , 
\end{cases}
\label{eq:R-sol-full}
\end{align}
where $B(y)$ is the solution to 
\begin{align}
\frac{B'(y)}{B(y)}
= \cot\bigl(h(y)\bigr) 
\label{eq:Bscaled-diffeq}
\end{align}
for $0 \le y<1$ with the initial condition $B(0)=1$. 
An alternative expression for $R_K(y)$ for $0 < y < 1$ is 
\begin{align}
R_K(y) = 
- e^{-iK y} 
+ \frac{2}{B(y)}
- \frac{2 i K }{B(y)}
\int_0^{y} {B}(y') \, e^{-iK y'} \, dy'
\ . 
\label{eq:R-sol-alt}
\end{align}
Using \eqref{eq:theta-scaled} and the smoothness of~$h$, 
it follows from \eqref{eq:Bscaled-diffeq} that $1/B(y)$ 
and all of its derivatives tend to zero as $y \to 1_-$, 
and this can be used to show from \eqref{eq:R-sol-alt} that 
$R_K(y)$ is a smooth function of $y$ everywhere, including $y=1$. 
It follows that $E_k(u)$ is a smooth function of~$u$. 

\begin{figure}[p]
\centering
\includegraphics[width=0.5\textwidth]{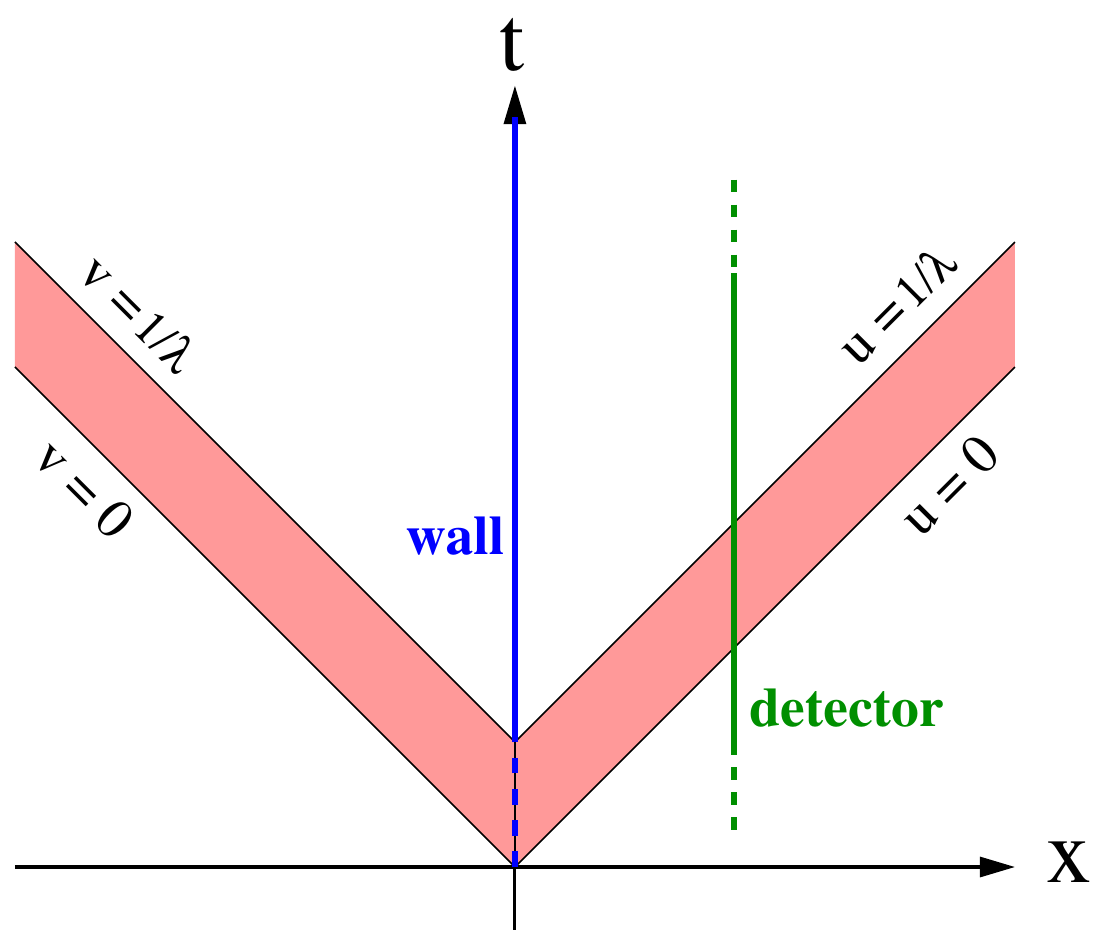}
% {\Huge Insert Minkowski figure here} 
\caption{$(1+1)$-dimensional Minkowski spacetime 
with a wall evolving at $x=0$. 
The wall starts to evolve at $(t,x) = (0,0)$ and becomes a 
fully-developed two-sided Dirichlet wall at 
$(t,x) = (\lambda^{-1},0)$. 
The wall sends a pulse of energy that travels to the right in the null strip 
$0 < u < \lambda^{-1}$ 
and to the left in the null strip $0 < v < \lambda^{-1}$.
The figure shows also the world line 
% \eqref{eq:detector-trajectory} 
of an inertial detector at $x=d>0$.}
\label{fig:minkowski}
\end{figure}

At $u \le 0$ and $u \ge \lambda^{-1}$, the mode functions $U_k$
reduce respectively to 
\begin{align}
U_k(u,v) = 
\begin{cases}
{\displaystyle{\frac{1}{\sqrt{8\pi k}} 
\left[
e^{-ikv} + e^{-iku}
\right]}}
& \text{for $u \le 0$}
\ , 
\\[2ex]
{\displaystyle{\frac{1}{\sqrt{8\pi k}} 
\left[
e^{-ikv} - e^{-iku}
\right]}}
& \text{for $u \ge \lambda^{-1}$}
\ . 
\end{cases}
\label{eq:U-sol-early-and-late}
\end{align}
At $u \le 0$, $U_k$ have not yet been affected by the wall, 
and they coincide with the usual spatially even mode functions in Minkowski, 
positive frequency with respect to~$\partial_t$. 
At $u \ge \lambda^{-1}$, $U_k$ feel the fully-developed Dirichlet wall, 
and they coincide with the half-space mode functions 
with the Dirichlet boundary condition. 
In the interpolating region, $0 < u < \lambda^{-1}$, 
$U_k$ are given by 
\eqref{eq:Umode-ansatz}
with 
\eqref{eq:E-intermsof-R}--\eqref{eq:Bscaled-diffeq}. 
The different regions are illustrated in Figure~\ref{fig:minkowski}. 

Recalling that the above formulas hold for $x>0$ and
the corresponding formulas for $x<0$ are obtained by spatial evenness, 
it can be verified that $U_k$
satisfy the usual Klein-Gordon orthonormality relations 
\begin{subequations}
\label{eq:U-KG-orthonormality}
\begin{align}
\bigl(U_k, U_{k'}\bigr) &= \delta(k-k') 
\ , 
\\
\bigl( \, \overline{U_k}, \overline{U_{k'}} \, \bigr) &= - \delta(k-k') 
\ , 
\\
\bigl(U_k, \overline{U_{k'}} \, \bigr) &= 0 
\ . 
\end{align}
\end{subequations}
where $(\,\cdot\,,\,\cdot\,)$ is 
the Klein-Gordon (indefinite) inner product~\cite{birrell-davies}.

\begin{figure}[p]
\centering
\includegraphics[width=0.5\textwidth]{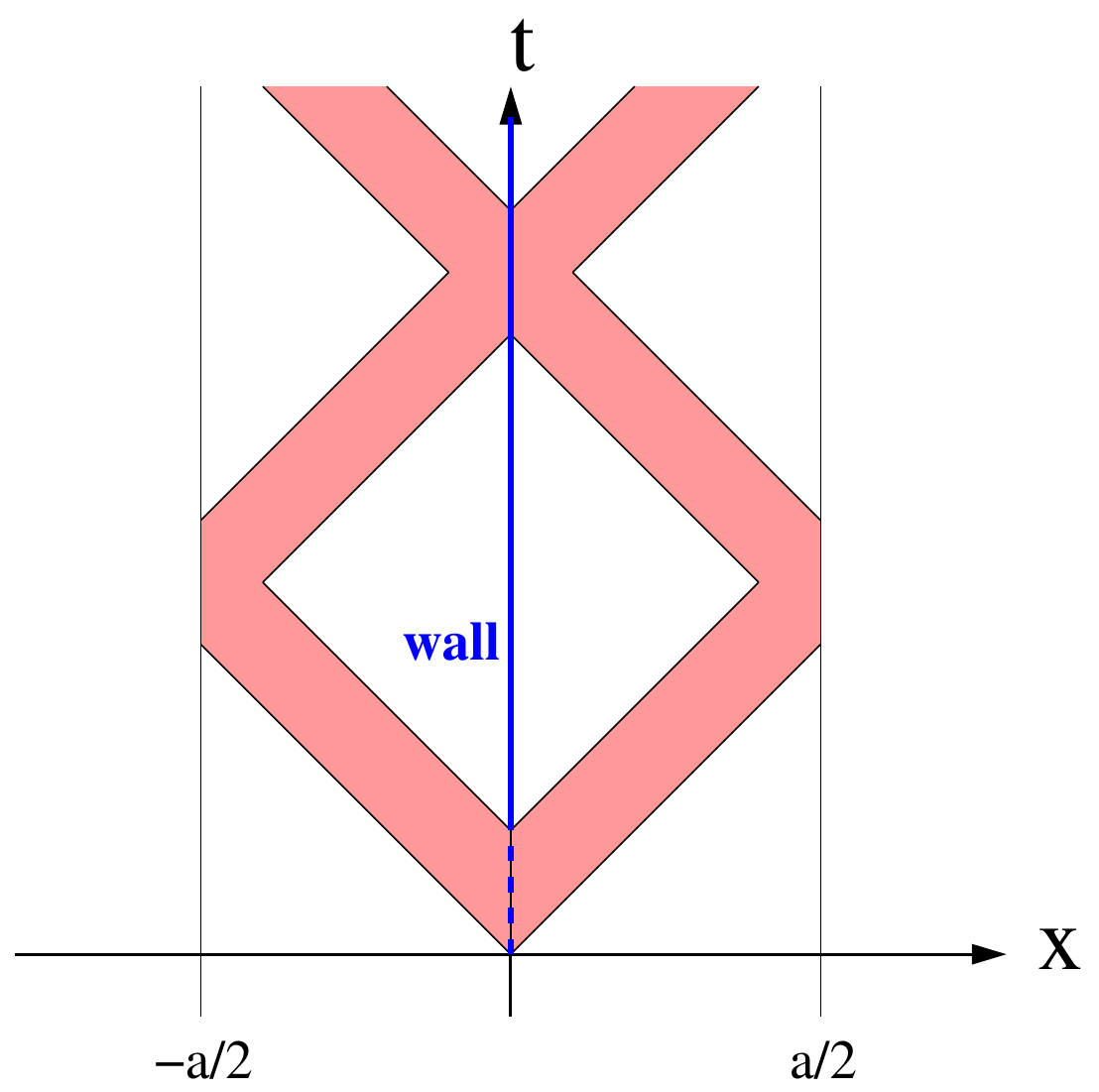}
% {\Huge Insert cavity figure here} 
\caption{$(1+1)$-dimensional Dirichlet cavity of length $a$ 
with a wall evolving at the centre, $x=0$. 
The wall evolution is as in Figure~\ref{fig:minkowski}, 
but the reflections from the boundaries at $x = \pm a/2$ 
affect the evolution of the mode functions for sufficiently late times. 
The figure shows the case $a >  2/\lambda$, 
in which the Dirichlet wall at $x=0$ has 
fully formed before the changes in 
the field due to the 
wall evolution reach the boundaries 
at $(t,x) = (a/2, \pm a/2)$.}
\label{fig:cavity}
\end{figure}

\subsection{Quantisation and the rapid wall creation limit}

We quantise the field in the usual fashion, 
adopting $U_k$ as the positive norm mode functions 
in the spatially even sector and 
the usual spatially odd Minkowski mode functions in the spatially odd sector. 
The spatially even part of $\phi$ is expanded as 
\begin{align}
\phi_{\text{even}} = \int_0^\infty 
\bigl(a_k U_k + a_k^\dagger \overline{U_k} \, \bigr) 
\, dk
\ , 
\label{eq:phi-even-expansion}
\end{align}
where the nonvanishing commutators of the annihilation and creation operators are 
$\bigl[a_k, a_{k'}^\dagger\bigr] = \delta(k-k')$. 
We denote by $|0_M\rangle$ the normalised state that is
annihilated by all $a_k$ and by all the usual Minkowski annihilation operators of 
the spatially odd sector. 
$|0_M\rangle$~is indistinguishable from the usual Minkowski vacuum 
in the region $t < |x|$ which is outside the causal future of the wall. 

We are interested in the energy that is transmitted   
into the quantum field by the evolving wall. Recall first that 
the classical stress-energy tensor of a massless minimally coupled scalar field 
is given by $T_{uu} = (\partial_u \phi) (\partial_u \phi)$,  
$T_{vv} = (\partial_v \phi) (\partial_v \phi)$
and $T_{uv} = 0$~\cite{birrell-davies}. 
We point-split the quantised versions of these expressions and express 
their expectation values in $|0_M\rangle$ in terms of the Wightman function of the field, using \eqref{eq:Umode-ansatz} 
and~\eqref{eq:phi-even-expansion}. Subtracting the Minkowski contribution and 
taking the coincidence limit, we find that 
the renormalised stress-energy tensor $\langle T_{ab} \rangle$ 
is given by 
\begin{subequations}
\label{eq:Tab-Minkowski}
\begin{align}
\langle T_{vv}\rangle &= \langle T_{uv}\rangle =0
\ , 
\\[1ex]
\langle T_{uu}\rangle &= \int_\mu^{\infty}
\frac{dk}{8\pi k}
\left(
{\bigl|E'_k(u)\bigr|}^2 - k^2
\right)
\ , 
\label{eq:Tuu-Minkowski}
\end{align}
\end{subequations}
where the constant $\mu$ is an infrared 
cutoff which we have inserted by hand. 

When $\mu>0$, $\langle T_{uu}\rangle$ 
is well defined for all~$u$, and vanishing for $u\le 0$ and
$u\ge \lambda^{-1}$, 
as is seen from \eqref{eq:E-intermsof-R} and~\eqref{eq:R-sol-full}. 
The convergence of the integral in \eqref{eq:Tuu-Minkowski} 
at $k\to\infty$ for $0 < u < \lambda^{-1}$ follows because 
${\bigl|E'_k(u)\bigr|}^2 = k^2 + O\bigl(k^{-2}\bigr)$ at large~$k$,  
as can be verified by repeated integration by parts in~\eqref{eq:R-sol-full}, 
integrating the exponential factor~\cite{wong}. 
When $\mu=0$, 
$\langle T_{uu}\rangle$ 
is still well defined and vanishing for 
$u\le 0$ and
$u\ge \lambda^{-1}$, but it is infrared divergent for 
$0 < u < \lambda^{-1}$: this follows because 
for $0 < u < \lambda^{-1}$ \eqref{eq:E-intermsof-R} and \eqref{eq:R-sol-alt} 
give ${|E'_k(u)|}^2 = 4 \lambda^2 
{[{B'}(\lambda u)]}^2 {[B(\lambda u)]}^{-4} + O(k^2)$ at small~$k$, 
and \eqref{eq:Bscaled-diffeq} shows that 
${B'}(\lambda u) {[B(\lambda u)]}^{-2}$ is nonvanishing. 
The infrared divergence was previously 
observed within a perturbative 
analysis in~\cite{Obadia:2001hj}. 

In words, this means that a positive infrared cutoff is 
required to make $\langle T_{ab}\rangle$ finite on the light
cone of each wall point where the wall has started to form but 
has not yet reached the Dirichlet form. 
Where $\langle T_{ab}\rangle$ is nonzero, 
it corresponds to null radiation travelling away from the wall. 

The total energy transmitted into the quantum field during the 
creation of the wall is 
\begin{align}
\langle E_{\text{tot}} \rangle 
& = 
\int_\Sigma \langle T_{tt} \rangle \, dx 
\notag
\\[1ex]
& = 
2 \int_0^{1/\lambda} \langle T_{uu} \rangle \, du  
\ , 
\label{eq:Ttot-Mink-def}
\end{align}
where for $\Sigma$ we may take any a constant $t$ 
hypersurface in the region $t > \lambda^{-1}$, 
and the last expression in \eqref{eq:Ttot-Mink-def} follows using 
\eqref{eq:Tab-Minkowski} 
and by including the contribution from $x<0$. 
Inserting the solution 
\eqref{eq:E-intermsof-R}--\eqref{eq:Bscaled-diffeq}
in \eqref{eq:Tab-Minkowski}, 
we find 
\begin{align}
\langle E_{\text{tot}} \rangle
= \frac{\lambda}{4\pi}
\int_{\mu/\lambda}^\infty \frac{dK}{K} 
\left(
\int_0^1{\bigl|R'_K(y)\bigr|}^2 \, dy \  - K^2
\right)
\ . 
\label{eq:Etot-Mink-scaled}
\end{align}

For rapid wall creation, we consider the limit of large $\lambda$ with fixed~$h$. 
Recall from \eqref{eq:R-sol-alt} that for $0 < y < 1$ we have 
${|R'_K(y)|}^2 = 4 
{[{B'}(y)]}^2 {[B(y)]}^{-4} + O(K^2)$, 
where the first term is bounded because 
$1/B(y)$ and its derivatives tend to zero as $y\to 1_-$. 
From \eqref{eq:Etot-Mink-scaled} we hence obtain 
\begin{align}
\langle E_{\text{tot}} \rangle
= \frac{\lambda}{\pi}
\! \left(
\ln(\lambda/\mu) \! \int_0^1 \frac{{[{B'}(y)]}^2}{{[B(y)]}^{4}} \, dy 
\  + O(1) \right) 
\ . 
\label{eq:Etot-Mink-as}
\end{align}

We conclude that in the rapid wall creation limit 
the energy transmitted into the quantum field 
diverges proportionally to $\lambda\ln(\lambda/\mu)$. 
The energy comes out as an increasingly narrow 
pulse near the light cone of the point $(t,x)=(0,0)$ but 
the magnitude of the pulse grows so rapidly 
that the stress-energy tensor does not have a 
distributional limit and the total energy diverges.

\section{Wall creation within a Dirichlet cavity\label{sec:cavity}}

In this section we adapt the analysis of Section 
\ref{sec:Minkowski}
to a wall that is created at the centre of a static cavity
whose left and right walls have time-independent Dirichlet boundary conditions. 
The main point of this adaptation is to verify that there 
is no need to introduce an infrared cutoff by hand since 
such a cutoff is already provided by the cavity.

\subsection{Classical field and mode functions}

Following the notation of Section~\ref{sec:Minkowski}, 
we confine the field $\phi$ to a static cavity whose walls are at $x= \pm a/2$,
where the positive constant $a$ is the length of the cavity. 
We take $\phi$ to satisfy the Dirichlet boundary condition at $x= \pm a/2$.

At the centre of the cavity, $x=0$, we introduce the time-dependent 
boundary condition as in Section~\ref{sec:Minkowski}, 
with the same assumptions about~$\theta(t)$.
Again, the boundary condition does not affect the
spatially odd part of the field, and it suffices to 
consider the spatially even part. 
We write down the formulas 
assuming $0 < x < a/2$, 
with the spatial evenness providing the formulas for $-a/2 < x<0$. 

We look for the mode solutions with the ansatz 
\begin{align}
V_n(u,v)  = \frac{1}{\sqrt{4\pi n}}
\bigl[
- F_n(v-a) + F_n(u)
\bigr]
\ , 
\label{eq:Vmode-ansatz}
\end{align}
where the index $n$ is an odd positive integer
and the function $F_n$ is to be found. 
This ansatz satisfies the wave equation at $0 < x < a/2$, 
and it satisfies $V_n(u,a+u)=0$, which is the Dirichlet boundary condition at 
$x=a/2$. 

Requiring 
\eqref{eq:Vmode-ansatz} to satisfy \eqref{eq:app:robin-plusside}
with $\theta=\theta(t)$ gives 
\begin{align}
L \sin\bigl(\theta(t)\bigr) 
\frac{d}{dt}
\left[
F_n(t-a) + F_n(t)
\right]
&= \cos\bigl(\theta(t)\bigr) 
\left[
F_n(t-a) - F_n(t)
\right]
\ . 
\label{eq:F-ev-eq}
\end{align}
We again parametrise $\theta=\theta(t)$ by~\eqref{eq:theta-param}. 
We choose the solution that for $u < \min(a, \lambda^{-1})$ is given by 
\begin{align}
F_n(u) = R_{\pi n {(\lambda a)}^{-1}}(\lambda u)
\ \ \text{for \ $u < \min(a, \lambda^{-1})$}
\ , 
\label{eq:F-intermsof-R}
\end{align}
where $R_K$ is given by 
\eqref{eq:R-sol-full} and~\eqref{eq:Bscaled-diffeq}. 
For $u\le0$ this implies 
\begin{align}
V_n(u,v) = 
\frac{1}{\sqrt{4\pi n}}
\bigl[
e^{-i \pi n v/a} + e^{-i \pi n u/a}
\bigr]
\ \ \ \text{for $u\le 0$}
\ ,  
\label{eq:Vn-early}
\end{align}
so that at early times $V_n$ are the
standard spatially even mode functions in the Dirichlet cavity. 
The domain $u < \min(a, \lambda^{-1})$, where the solution 
\eqref{eq:F-intermsof-R} holds, is where the time-dependence 
due to the evolving wall has not yet come back to $x=0$ 
after being reflected from $x=a/2$. 

To evolve $F_n$ further to the future, 
one needs to account for the reflections of the time-dependence 
that start to arrive to $x=0$. 
The case of main interest for us is when $\lambda > a^{-1}$, 
which occurs when $a$ is considered fixed and we consider 
a rapid wall formation. In this case 
the Dirichlet wall at $x=0$ 
is fully formed when the first reflection 
due to the wall evolution arrives back to $x=0$. 
Equation \eqref{eq:F-intermsof-R} then holds for 
$u < \lambda^{-1}$, so that 
$F_n(u) =  - e^{-i\pi n u/a}$ for $\lambda^{-1} \le u \le a$, 
and the evolution of $F_n(u)$ to $u>a$ is given just by 
successive Dirichlet reflections from $x=0$ and $x=a/2$. 
The case in which $\lambda > 2/a$ is illustrated 
in Figure~\ref{fig:cavity}.

\subsection{Quantisation and the rapid wall creation limit}

We again quantise the field in the usual fashion 
and denote by $|0_c\rangle$ the vacuum 
with the above choice for the above positive norm mode functions. 
$|0_c\rangle$ is indistinguishable from the usual Dirichlet cavity vacuum
in the region $t < |x|$, where its renormalised stress-energy tensor 
has the expectation value \cite{birrell-davies}
\begin{subequations}
\label{eq:Tab-cav-early}
\begin{align}
& \langle T_{uu}\rangle_{\text{(early)}} 
= \langle T_{vv}\rangle_{\text{(early)}} 
= - \frac{\pi}{96 \, a^2}
\ , 
\\[1ex]
& \langle T_{uv}\rangle_{\text{(early)}}  =0
\ . 
\end{align}
\end{subequations}

To examine the stress-energy tensor due to the wall creation, 
we assume $\lambda > 2/a$, 
and we consider the region $0 < x < a/2$ and $t < a/2$, 
as illustrated 
in Figure~\ref{fig:cavity}.
In this region the solution \eqref{eq:F-intermsof-R} holds, 
and the $v$-dependent part of $V_n$ has still the standard 
form proportional to~$e^{-inv/a}$. 
Writing 
\begin{align}
\langle T_{ab}\rangle = \langle T_{ab}\rangle_{\text{(early)}} 
+ \Delta \langle T_{ab}\rangle
\ ,  
\label{eq:Tab-decomp}
\end{align}
we find 
\begin{subequations}
\label{eq:DeltaTab-cavity}
\begin{align}
\Delta \langle T_{vv}\rangle & = \Delta \langle T_{uv}\rangle =0
\ , 
\\[1ex]
\Delta \langle T_{uu}\rangle
&= 
\sum_{n>0\ \text{odd}}
\frac{1}{4\pi n} 
\left[
{\bigl|F'_n(u)\bigr|}^2 - {(\pi n/a)}^2
\right]
\ , 
\label{eq:DeltaTuu-cavity}
\end{align}
\end{subequations}
where the convergence of the sum 
in \eqref{eq:DeltaTuu-cavity}
at large $n$ can be verified as in Section~\ref{sec:Minkowski}, 
and there is no infrared divergence because the sum starts at $n=1$. 
$\Delta \langle T_{uu}\rangle$ is vanishing for 
$u \le0$ and for $u\ge \lambda^{-1}$. 

The total energy transmitted into the quantum field is given as in 
\eqref{eq:Ttot-Mink-def}
but with $\langle T_{ab}\rangle$ replaced by 
$\Delta \langle T_{ab}\rangle$, and 
$\Sigma$ being now any 
constant $t$ hypersurface at $\lambda^{-1} < t < a/2$. Using 
\eqref{eq:DeltaTuu-cavity} with~\eqref{eq:F-intermsof-R}, we obtain 
\begin{align}
\langle E_{\text{tot}} \rangle
= \frac{\lambda}{2\pi}
\sum_{n>0\ \text{odd}}
\frac{1}{n}\left[
\int_0^1{\bigl|R'_{\pi n {(\lambda a)}^{-1}}(y)\bigr|}^2 \, dy \  
- \left(\frac{\pi n}{\lambda a}\right)^2
\right]
\ . 
\label{eq:Etot-cavity-scaled}
\end{align}
In the limit of large $\lambda$, we may approximate the sum by an integral,
and using the properties of $R_K$ as in Section~\ref{sec:Minkowski} gives 
\begin{align}
\langle E_{\text{tot}} \rangle
= \frac{\lambda}{\pi}
\! \left[
\ln \! \left(\frac{\lambda a}{\pi}\right) 
\int_0^1 \frac{{[{B'}(y)]}^2}{{[B(y)]}^{4}} \, dy 
\ + O(1) \right]
\ . 
\label{eq:Etot-cavity-as}
\end{align}
The energy diverges proportionally to $\lambda \ln(\lambda a/\pi)$, 
and comparison with \eqref{eq:Etot-Mink-as} shows that 
$\pi/a$ plays the role of an infrared cutoff. 
The divergence implies that the stress-energy tensor does not 
have a distributional limit at $\lambda\to\infty$.

\section{Wall creation in Minkowski space over 
infinite time\label{sec:example}}

In this section we adapt the Minkowski space analysis of 
Section \ref{sec:Minkowski} 
to a specific one-parameter family of wall evolution profiles
for which the evolution is 
nontrivial at all finite times but 
reduces to no wall in the asymptotic past and to 
a wall with nonvanishing reflection and 
transmission coefficients in the asymptotic future. 
The main point is to verify that passing to an 
appropriate limit within this one-parameter family  
allows us again to model a rapid creation of a Dirichlet wall, 
and the results for the stress-energy tensor agree 
with those in Section~\ref{sec:Minkowski}. 
These properties will justify our use of this one-parameter 
family of evolution profiles with 
a particle detector in Section~\ref{sec:detector-on-Minkowski}. 

We take the boundary condition to be as in 
\eqref{eq:scaled-BC} 
with $\lambda$ a positive parameter and 
\begin{align}
h(y) = \arctan(1+e^{-y})
\ , 
\label{eq:h-incomplete}
\end{align}
so that 
\begin{align}
\theta(t) = \arctan \! \left(\frac{1 + e^{-\lambda t}}{\lambda L}\right)
\ . 
\label{eq:infinite-profile}
\end{align}
Since $0 < \theta(t) < \pi/2$, the wall exists for all~$t$, 
and it is never Dirichlet. 
Since $\theta(t) \to \pi/2$ as $t\to-\infty$, 
the wall disappears in the asymptotic past, 
and the wall formation starts exponentially slowly. 
Since $\theta(t) \to \arccot(\lambda L)$ as $t\to\infty$, 
the end state of the wall in the asymptotic future is not Dirichlet, 
but it can be made arbitrarily close to Dirichlet by taking 
$\lambda L$ large. 

The parameter $\lambda$ has hence a dual role: 
it determines both how rapid the wall formation is 
and how close the wall is to Dirichlet in the asymptotic future. 
In the limit $\lambda\to\infty$, 
we approach the instantaneous creation of a 
Dirichlet wall at $t=0$. 

We proceed as in Section \ref{sec:Minkowski}. 
Equation \eqref{eq:R-sol-full} is now replaced by 
\begin{align}
R_K(y) = 
e^{-iK y} - \frac{2}{B(y)}
\int_{-\infty}^{y} {B'}(y') \, e^{-iK y'} \, dy'
\ , 
\label{eq:R-sol-infty}
\end{align}
where \eqref{eq:Bscaled-diffeq}
and the initial condition $B(y)\to1$ as $y\to-\infty$ give 
\begin{align}
B(y) = 1 + e^y
\ . 
\label{eq:B-example}
\end{align}
We find that $U_k$ is given by 
\eqref{eq:Umode-ansatz}
with 
\begin{align}
E_k(u) = \frac{e^{-iku}}{1+e^{\lambda u}}
\left[
1 - \left(\frac{\lambda + ik}{\lambda -ik} \right)e^{\lambda u}
\right]
\ . 
\label{eq:Esol-sampleprof}
\end{align}

For the stress-energy tensor, \eqref{eq:Tab-Minkowski} gives 
\begin{subequations}
\begin{align}
\langle T_{vv}\rangle &= \langle T_{uv}\rangle =0
\ , 
\\[1ex]
\langle T_{uu}\rangle &= 
\frac{\lambda^2}{32 \pi \cosh^4(\lambda u/2)}
\int_\mu^\infty 
\frac{dk}{k \left[1 + {(k/\lambda)}^2\right]}
\notag
\\[1ex]
&= 
\frac{\lambda^2 \ln\!\left[1 + {(\lambda/\mu)}^2\right]}{64\pi \cosh^4(\lambda u/2)}
\ , 
\label{eq:Tuu-sample-Minkowski}
\end{align}
\end{subequations}
where the positive infrared cutoff $\mu$ is again needed 
to make $\langle T_{uu}\rangle$ finite. 

When $\lambda\to\infty$, $\langle T_{uu}\rangle$ 
vanishes for $u\ne0$ and diverges for $u=0$. 
To examine the strength of this divergence,  
we write 
\begin{subequations}
\label{eq:Tuu-sample-redMinkowski}
\begin{align}
\langle T_{uu}\rangle &= 
\frac{\lambda \ln\!\left[1 + {(\lambda/\mu)}^2\right]}{24\pi} 
\, f_\lambda(u)
\ , 
\\[1ex]
f_\lambda(u)
&= \frac{3\lambda}{8\cosh^4(\lambda u/2)}
\ , 
\end{align}
\end{subequations}
and observe that $f_\lambda(u) \to \delta(u)$ as $\lambda\to\infty$. 
The divergence is hence too strong for $\langle T_{uu}\rangle$ 
to have a distributional limit. 
The total energy transmitted into the quantum field is
\begin{align}
\langle E_{\text{tot}} \rangle 
& = 
\lim_{t\to\infty}\int_{\Sigma_t} 
\langle T_{tt} \rangle \, dx 
\notag
\\[1ex]
& = 
\lim_{t\to\infty} 
2 \int_{-\infty}^{t} \langle T_{uu} \rangle \, du  
\notag
\\[1ex]
& = 
\frac{\lambda \ln\!\left[1 + {(\lambda/\mu)}^2\right]}{12\pi} 
\ , 
\label{eq:Etot-Mink-infinite}
\end{align}
where $\Sigma_t$ is a hypersurface at constant~$t$, 
and the final expression comes using \eqref{eq:Tuu-sample-redMinkowski} 
and observing that $\int_{-\infty}^\infty f_\lambda(u) \, du = 1$. 
In the limit $\lambda\to\infty$, 
% $\langle E_{\text{tot}} \rangle $ 
the energy diverges proportionally to 
${(6\pi)}^{-1}\lambda\ln(\lambda/\mu)$ 
and comes out as a narrow burst
near the light cone of $(t,x)=(0,0)$.

\section{Response of an Unruh-DeWitt detector 
to rapid wall creation\label{sec:detector-on-Minkowski}} 

In this section we consider the response of an inertial 
Unruh-DeWitt particle detector to the creation of a wall. 
We work in Minkowski spacetime with the wall 
creation profile~\eqref{eq:infinite-profile}. 
We are interested in the limit of large~$\lambda$, 
in which the burst of energy from the wall diverges 
on the light cone of $(t,x)=(0,0)$. 
We ask what happens in the limit of large $\lambda$ 
to the response of a detector that crosses this light cone.

\subsection{Detector and its trajectory}

We consider a version of the Unruh-DeWitt detector \cite{unruh,dewitt}
that couples linearly to the proper time derivative of the field
\cite{Louko:2014aba,Raine:1991kc,Raval:1995mb,Davies:2002bg,Wang:2013lex,Martin-Martinez:2014qda,Juarez-Aubry:2014jba,Hotta:2015yla}. 
Following the notation of~\cite{Juarez-Aubry:2014jba}, 
we denote by $\mathsf{x}(\tau)$ the detector's worldline, 
parametrised by the proper time~$\tau$. 
We assume that the coupling to the field is 
proportional to a real-valued function $\chi(\tau)$ 
that specifies how the interaction is turned on and off. 
We call $\chi$ the switching function and assume it 
to be smooth with compact support.

In first-order perturbation theory, 
the detector's probability to make a transition from a 
state with energy $0$ to a state with energy $\omega$ 
is proportional to the response function, given by 
\begin{align}
\mathcal{F}^{(1)}(\omega)
&=
\int^{\infty}_{-\infty}\,d\tau'\,\int^{\infty}_{-\infty}\,d\tau''\, 
e^{-i \omega(\tau'-\tau'')} \,\chi(\tau')\chi(\tau'') \, 
\partial_{\tau'}
\partial_{\tau''}
\mathcal{W}(\tau',\tau'')
\ , 
\label{eq:respfunc-def}
\end{align}
where the correlation function $\mathcal{W}$ 
is the pull-back of the Wightman 
function to the detector's worldline, 
\begin{align}
\mathcal{W}(\tau',\tau'') :=
\langle\psi|\phi\bigl(\mathsf{x}(\tau')\bigr)
\phi\bigl(\mathsf{x}(\tau'')\bigr)|\psi\rangle
\ , 
\label{eq:W-def}
\end{align}
and $|\psi\rangle$ is the state to which 
the field was initially prepared. 
The superscript ${}^{(1)}$ in \eqref{eq:respfunc-def}
is a reminder that the detector couples to the (first)  
derivative of the field. The derivatives in 
\eqref{eq:respfunc-def} are understood in the distributional sense, 
and integration by parts gives the alternative expression 
\begin{align}
\mathcal{F}^{(1)}(\omega)
&= 
\int_{-\infty}^\infty d\tau' \int_{-\infty}^\infty d\tau'' \, 
Q_\omega'(\tau') \overline{Q_\omega'(\tau'')} \, 
\mathcal{W}(\tau',\tau'')
\ , 
\label{eq:respfunc-altdef}
\end{align}
where $Q_\omega(\tau) := e^{-i\omega\tau}\chi(\tau)$. 
$\mathcal{F}^{(1)}$ is hence well defined whenever 
$\mathcal{W}$ is a well-defined distribution. 

We take the detector's trajectory to be
\begin{align}
(t,x) = (\tau+d,d)
\ , 
\label{eq:detector-trajectory}
\end{align}
where $d$ is a positive constant.
The detector is inertial and it crosses the light 
cone of the origin at $(t,x) = (d,d)$. 
The zero of the proper time has been chosen to occur at this crossing. 
The geometry is shown in Figure~\ref{fig:minkowski}.

\subsection{Preliminaries: Minkowski vacuum and 
Dirichlet half-space\label{subsec:static-comparisons}}

For comparisons to be made below, 
we record here the response in Minkowski vacuum 
and in Minkowski half-space with the Dirichlet boundary condition. 

When there is no wall and the field is in the usual Minkowski vacuum, 
the response function is given by \cite{Louko:2014aba} 
\begin{align}
\mathcal{F}^{(1)}_{\text{Mink}}(\omega)
&=
-\omega \Theta(-\omega) 
\int_{-\infty}^{\infty} d u \, {[\chi(u)]}^2
\notag
\\[1ex]
&\hspace{3ex}
+ 
\frac{1}{\pi}
\int^{\infty}_{0} 
ds \, 
\frac{\cos(\omega s)}{s^2} 
\int_{-\infty}^{\infty} d u \, 
\chi(u) [\chi(u) - \chi(u-s)] 
\ . 
\label{eq:F1-0M-inert}
\end{align}
$\mathcal{F}^{(1)}_{\text{Mink}}$ is 
independent of the infrared cutoff, and its asymptotic form at large 
$|\omega|$ is \cite{Louko:2014aba} 
\begin{align}
\mathcal{F}^{(1)}_{\text{Mink}}(\omega)
&=
-\omega \Theta(-\omega) 
\int_{-\infty}^{\infty} d u \, {[\chi(u)]}^2
\ + O^{\infty} \bigl(\omega^{-1}\bigr)
\ . 
\label{eq:F1-0M-inert-large}
\end{align}

When there is a static wall at $x=0$ 
and the field is in the usual vacuum state with Dirichlet conditions 
at this wall, we show in Appendix \ref{app:statichalfspace} 
that the response function is
\begin{align}
\mathcal{F}^{(1)}_{\text{Dir}}
= 
\mathcal{F}^{(1)}_{\text{Mink}}
+ 
\Delta_{\text{Dir}} \mathcal{F}^{(1)}
\ , 
\label{eq:DirF1-decomp}
\end{align}
where 
\begin{subequations}
\label{eq:deltaDirF1-finalwithG}
\begin{align}
\Delta_{\text{Dir}} \mathcal{F}^{(1)}(\omega)
& = 
\frac{1}{2\pi} 
\Realpart \Biggl\{ e^{-2i\omega d}
\Biggl[
i \pi G_\omega(2d)
\notag
\\[1ex]
& \hspace{13ex}
+ 
\int_0^\infty ds \, 
\frac{e^{-i\omega s} G_\omega(2d+s) - e^{i\omega s} G_\omega(2d-s)}{s}
\ \Biggr] \Biggr\}
\ , 
\label{eq:deltaDirF1-final}
\\
%\end{align}
%where 
%\begin{align}
G_\omega(y) &= 
\int_{-\infty}^\infty du \, 
\bigl[\chi'(u) - i \omega \chi(u) \bigr]
\chi(u - y)
\ ,  
\label{eq:G-def}
\end{align}
\end{subequations}
which is again independent of the infrared cutoff. 
We also show that the asymptotic large $|\omega|$ form of 
$\Delta_{\text{Dir}} \mathcal{F}^{(1)}$ is 
\begin{align}
\Delta_{\text{Dir}} \mathcal{F}^{(1)}(\omega)
& = 
\Theta(-\omega) 
\left[
\omega \cos(2d\omega) \int_{-\infty}^\infty du \, \chi(u) \chi(u - 2d) 
\right.
\notag
\\[1ex]
& \hspace{12ex}
\left. 
+ \sin(2d\omega) \int_{-\infty}^\infty du \, \chi'(u) \chi(u - 2d) 
\right]
%\notag
%\\[1ex]
%& \hspace{3ex}
\ + O^{\infty}\bigl(\omega^{-1}\bigr)
\ . 
\label{eq:DeltaDirF1-as}
\end{align}

\subsection{Evolving wall}

When the wall is present with the profile~\eqref{eq:infinite-profile}, 
we write the response function as 
\begin{align}
\mathcal{F}^{(1)}_\lambda 
= 
\mathcal{F}^{(1)}_{\text{Mink}}
+ 
\Delta \mathcal{F}^{(1)}_\lambda 
\ . 
\label{eq:F1lambda-decomp}
\end{align}
We show in Appendix \ref{app:lambdalimit} that 
$\Delta \mathcal{F}^{(1)}_\lambda$ has a finite 
limit as $\lambda\to\infty$, given by  
\begin{align}
\Delta \mathcal{F}^{(1)}(\omega)
&=
\int^{\infty}_{-\infty}\,d\tau'\,\int^{\infty}_{-\infty}\,d\tau''\, 
Q_\omega'(\tau') \overline{Q_\omega'(\tau'')} \, 
\Delta \mathcal{W}(\tau',\tau'')
\ , 
\label{eq:DeltaF-def}
\end{align}
where $\Delta \mathcal{W}(\tau',\tau'')$ is given by the following expressions: 
\begin{subequations}
\label{eq:DeltaW-def}
\begin{align}
\text{$\tau'>0$, $\tau''>0$}: &
\ \ 
- \frac{1}{4\pi}
\Bigl[ E_1\bigl(\epsilon + i \mu (\tau'-\tau'' - 2d)\bigr) 
+ E_1\bigl(\epsilon + i \mu (\tau'-\tau'' + 2d) \bigr) \Bigr]
\ , 
\\[1ex]
\text{$\tau'>0$, $\tau''<0$}: &
\ \ 
- \frac{1}{4\pi}
\Bigl[
E_1\bigl(\epsilon + i \mu (\tau'-\tau'' - 2d) \bigr) 
+ E_1\bigl(\epsilon + i \mu (\tau'-\tau'') \bigr) \Bigr]
\ , 
\\[1ex]
\text{$\tau'<0$, $\tau''>0$}: &
\ \ 
- \frac{1}{4\pi}
\Bigl[
E_1\bigl(\epsilon + i \mu (\tau'-\tau'') \bigr) 
+ E_1\bigl(\epsilon + i \mu (\tau'-\tau'' + 2d) \bigr) \Bigr]
\ , 
\\[1ex]
\text{otherwise}: &
\ \ \ 
0 
\ . 
\end{align}
\end{subequations}
Here $\mu$ is the infrared cutoff and is assumed positive.  
$E_1$~is the exponential integral in
the notation of~\cite{dlmf}, 
taking values on its principal branch 
in the sense of $\epsilon\to0_+$. 
We further show in Appendix \ref{app:lambdalimit} that when $\omega+\mu\ne0$, 
$\Delta \mathcal{F}^{(1)}$ 
can be put in the form 
\begin{subequations}
\label{eq:deltaF1-finalwithH}
\begin{align}
\Delta \mathcal{F}^{(1)}(\omega)
& = 
\frac{{[\chi(0)]}^2}{2\pi}
\ln\bigl|1 + (\omega/\mu)\bigr|
\notag
\\[1ex]
& \hspace{3ex}
- 
\frac{\omega}{2\pi} \int_0^\infty 
ds \, \frac{\sin\bigl((\omega+\mu)s\bigr)}{s} 
\int_0^s du \, \chi(u) \chi(u-s) 
\notag
\\[1ex]
& \hspace{3ex}
+ 
\frac{1}{2\pi} \int_0^\infty
ds \, \frac{\cos\bigl((\omega+\mu)s\bigr)}{s} 
\left(
\chi(0) \bigl[\chi(0) - \chi(-s)\bigr] - \int_0^s du \, \chi(u) \chi'(u-s) 
\right)
\notag
\\[1ex]
& \hspace{3ex}
+ \frac{1}{2\pi} 
\Realpart \Biggl\{ e^{-2i\omega d}
\Biggl[
i \pi H_\omega(2d)
\notag
\\[1ex]
& \hspace{15ex}
+ 
\int_0^\infty ds \, 
\frac{e^{-i(\mu+\omega) s} H_\omega(2d+s) - e^{i(\mu+\omega) s} H_\omega(2d-s)}{s}
\ \Biggr] \Biggr\}
\ , 
\label{eq:deltaF1-final}
\\
%\end{align}
%where 
%\begin{align}
H_\omega(y) &= 
\int_0^\infty du \, 
\bigl[\chi'(u) - i \omega \chi(u) \bigr]
\chi(u - y)
\ . 
\label{eq:H-def}
\end{align}
\end{subequations}

Four observations are in order. 

First, given that $\mu$ is assumed positive, 
equations 
\eqref{eq:DeltaF-def}
and \eqref{eq:DeltaW-def} show that 
$\Delta \mathcal{F}^{(1)}$ is manifestly finite. 
The detector's response remains finite when the wall 
creation becomes instantaneous, even though the detector passes 
through an infinite pulse of energy. 

Second, $\Delta \mathcal{F}^{(1)}$ has a finite $\mu\to0$ limit 
if and only if $\chi(0)=0$. This is seen from  
\eqref{eq:deltaF1-final} where the only potential divergence
at $\mu\to0$ comes from the first term. 
The infrared cutoff can hence be removed if and only 
if the detector does not operate at the moment of crossing
the light cone of the wall creation event. 

Third, as a consistency check, we note that if $\chi(\tau)$ 
vanishes for $\tau\le0$, the first three terms in \eqref{eq:deltaF1-final}
vanish, and comparison of 
\eqref{eq:deltaF1-finalwithH}
and 
\eqref{eq:deltaDirF1-finalwithG} 
shows that
$\Delta \mathcal{F}^{(1)}$ reduces to 
$\Delta_{\text{Dir}} \mathcal{F}^{(1)}$ 
if $\mu$ is taken to zero. 
If the detector operates only after crossing 
the light cone of the wall creation event, 
the response is identical to that in a half-space 
with a static Dirichlet wall. 

Fourth, we verify in Appendix \ref{app:lambdalimit} that the asymptotic form of 
$\Delta \mathcal{F}^{(1)}$ at large energy gap is 
\begin{align}
\Delta \mathcal{F}^{(1)}(\omega)
& = 
\frac{{[\chi(0)]}^2}{2\pi}
\ln\bigl( e^{-1} \bigl|1 + (\omega/\mu)\bigr| \bigr) 
\notag
\\[1ex]
& \hspace{3ex}
+ 
\Theta(-\omega-\mu) 
\left[
\omega \cos(2d\omega) \int_0^\infty du \, \chi(u) \chi(u - 2d) \right.
\notag
\\[1ex]
& \hspace{19ex}
\left. 
+ \sin(2d\omega) \int_0^\infty du \, \chi'(u) \chi(u - 2d) 
\right]
%\notag
%\\[1ex]
%& \hspace{3ex}
\ + O\bigl(\omega^{-1}\bigr)
\ . 
\label{eq:DeltaF1-as}
\end{align}
The terms proportional to 
$\omega \cos(2d\omega)$ and $\sin(2d\omega)$ are as expected 
from the corresponding terms in 
$\Delta_{\text{Dir}} \mathcal{F}^{(1)}$Z~\eqref{eq:DeltaDirF1-as}. 
The additional term, proportional to ${[\chi(0)]}^2$, comes strictly 
from the moment of  
crossing the light cone of the wall creation event. 
This term is dominant for 
$\omega\to\infty$ and subdominant for $\omega\to-\infty$.

\section{Discussion\label{Discussion}}

The purpose of this work has been to 
present a formalism 
for discussing the smooth creation of boundary conditions in quantum field theory, 
and to highlight some preliminary findings of interest. Specifically, we have examined several properties of the 
energy flux resulting from the smooth creation of a Dirichlet boundary condition for a massless 
scalar field in flat $(1+1)$-dimensional spacetime, and the resulting response of a particle detector. We have paid particular attention to the sharp creation limit of such a procedure. This type of scenario has gained interest recently from a number of different perspectives, and is markedly different from the more standard setting of a moving boundary condition. Our primary findings from this work are the following.

First, we have
shown that the creation of a wall in Minkowski space  
induces an energy flux that is infrared divergent, 
regardless how slowly and smoothly the creation unfolds. 
This divergence was previously observed within a 
perturbative analysis in~\cite{Obadia:2001hj}, and our results 
confirm that the divergence transcends the perturbative framework.  
While the Wightman function of the $(1+1)$-dimensional massless field is well known to be infrared divergent, it may be surprising that in our situation the infrared divergence shows up also in the stress-energy expectation value, which involves the Wightman function only through its derivatives. The upshot seems to be that in our time-dependent situation the infrared divergence of the Wightman function can no longer be thought of as an infinite additive constant but must be regarded as an infinite function, which does not drop out on taking a derivative. It should be interesting to give this phenomenon a more precise mathematical description, especially given its surprising and unintuitive nature.

Second, we have demonstrated that in the sharp creation limit 
(i.e.\ instantaneously producing a mirror) the resulting energy density flux 
is UV divergent, and diverges stronger than in any distributional sense. 
Thus, such a process would input an infinite amount of energy into the field. 
Indeed such a result is to be 
expected~\cite{unruh-seoultalk,Asplund:2013zba,Anderson:1986ww,trousers-revisited,Obadia:2001hj}, 
and as demonstrated in \cite{Brown:2014qna} is related to the 
fact that the entanglement entropy between the two 
regions on either side of the created wall is UV divergent.

Third, we have considered the response of an inertial derivative-coupling Unruh-DeWitt detector that crosses the energy flux emitted from the wall creation. We showed that the detector's response remains finite in the limit of instantaneous wall creation, despite the infinite amount of energy that the sharp creation injects into into the field. We also showed that in this sharp wall creation limit the detector's response depends on the infrared cutoff, even though the derivative-coupling detector is known to be insensitive to the infrared ambiguity of the Wightman function in a number of other quantum states. Both of these properties are similar to the response of an inertial detector in a 
Minkowski spacetime model \cite{Louko:2014aba,Martin-Martinez:2015dja} of a black hole firewall~\cite{Almheiri:2012rt}, 
and they add to the evidence that the prospective ability of a 
black hole firewall to resolve the black hole information paradox 
must hinge on the firewall's detailed
gravitational structure. 

Our detector results were obtained in Section \ref{sec:detector-on-Minkowski} 
under a specific one-parameter family of wall creation profiles. 
We conjecture that the same results for the sharp creation limit ensue 
within the full family of profiles introduced in Section~\ref{sec:Minkowski}. 
It is straightforward to verify that within this full family 
the pointwise sharp creation limit of the Wightman function 
is still given by~\eqref{eq:DeltaW-def}; 
to justify the conjecture, 
it would remain to show that the sharp creation limit in the response function 
\eqref{eq:respfunc-altdef}
can be taken pointwise under the integral. This question warrants further consideration. 

An interesting next step would be to examine the entanglement 
structure between the bursts of 
particles generated by smooth wall creation, 
with the aim of showing how the formalism and results of 
\cite{Brown:2014qna} emerge in the sharp creation limit 
and comparing with the conformal field 
theory treatments of~\cite{Asplund:2013zba,Asplund:2014coa}. 
As preparation for this analysis, we give in 
Appendix \ref{app:bogocoeffs} the Bogoliubov coefficients between the 
field modes adapted to the boundary condition 
before and after the creation of the wall. 
Another next step would be to examine how this entanglement 
may be harvested by particle detectors. Conversely, it would be interesting to examine how 
pre-existing entanglement between particle detectors 
is affected by the wall creation, in the formalism that 
was applied to a Rindler firewall in~\cite{Martin-Martinez:2015dja}.

Finally, we have throughout maintained that the quantum 
field lives on a nondynamical 
Minkowski metric even when the energy in the quantum field became infinite. 
Allowing the metric to become dynamical and to respond to the growing 
stress-energy could provide a model for a firewall in an 
evaporating black hole spacetime, in which the gravitational aspects near the horizon have 
had time to become significant.

\section*{Acknowledgments}

We thank Jason Doukas, 
Chris Fewster, Keijo Kajantie, 
Esko Keski-Vakkuri
and 
Bill Unruh 
for helpful 
discussions and 
correspondence, 
and an anonymous referee for helpful suggestions. 
E.B. acknowledges support by the Michael 
Smith Foreign Study Supplements Program. 
J.L. was supported in part by STFC (Theory
Consolidated Grant ST/J000388/1).

% \newpage 

\begin{appendix}

\section{$-\partial_x^2$ on a line with 
a distinguished point\label{app:sa-linewithpoint}}

In this appendix we collect relevant properties 
about the self-adjoint extensions of the operator 
$-\partial_x^2$ on $L_2(\BbbR \setminus\{0\})$. 
The general theory can be found for example in 
\cite{reed-simonII,blabk} and a pedagogical summary in~\cite{Bonneau:1999zq}. 

We take the coordinate $x$ to have the physical dimension of length. 
The self-adjoint extensions of $-\partial_x^2$ form a $U(2)$ family, 
specified by the boundary condition 
\cite{Bonneau:1999zq}
\begin{align}
\begin{pmatrix}
L \psi_+' - i \psi_+ \\
L \psi_-' + i \psi_-\\
\end{pmatrix} 
= 
U 
\begin{pmatrix}
L \psi_+' + i \psi_+ \\
L \psi_-' - i \psi_-\\
\end{pmatrix} 
\ , 
\label{eq:app:sa-gen}
\end{align}
where $\psi$ is the (generalised) eigenfunction, 
$\psi_\pm := \lim_{x\to0_\pm} \psi(x)$, 
$\psi'_\pm := \lim_{x\to0_\pm} \psi'(x)$, 
$L$ is a positive constant of dimension length and $U\in U(2)$. 
The constant $L$ has been introduced for dimensional convenience 
and its value is considered fixed. 
The extensions are then uniquely parametrised by the matrix~$U\in U(2)$. 
Physically, $U$ encodes the reflection and transmission 
coefficients across $x=0$. 

We specialise to the one-parameter subgroup of $U(2)$ given by 
\begin{align}
U(\theta) = e^{-i\theta} 
\begin{pmatrix}
\cos\theta & i \sin\theta \\
i \sin\theta & \cos\theta \\
\end{pmatrix} 
\ , 
\ \ \ 
\theta \in [0, \pi)
\ , 
\label{eq:app:U-theta}
\end{align}
and we denote the corresponding self-adjoint 
extensions of $-\partial_x^2$ by $- \Delta_\theta$. 
If $\psi_+=\psi_-=0$ and $\psi'_+=\psi'_-$, 
\eqref{eq:app:sa-gen} is satisfied as an identity. 
If $\psi_+=\psi_-$ and $\psi'_+=-\psi'_-$, \eqref{eq:app:sa-gen} becomes 
\begin{subequations}
\label{eq:app:robin}
\begin{align}
L \psi_+' \sin\theta &= \psi_+ \cos\theta
\ , 
\label{eq:app:robin-plusside}
\\
- L \psi_-' \sin\theta &= \psi_- \cos\theta
\ . 
\end{align}
\end{subequations}
$- \Delta_\theta$ hence leaves the even and 
odd subspaces of $L_2(\BbbR \setminus\{0\})$ invariant. 

On the odd subspace of $L_2(\BbbR \setminus\{0\})$, 
$- \Delta_\theta$ reduces to 
the unique self-adjoint extension of $-\partial_x^2$ 
on the odd subspace of $L_2(\BbbR)$. 
The generalised eigenfunctions are proportional 
to $\sin(kx)$ where $k>0$, and the spectrum is the positive continuum. 

On the even subspace of $L_2(\BbbR \setminus\{0\})$, 
$- \Delta_\theta$ is determined by 
the Robin boundary condition \eqref{eq:app:robin} 
on each side of $x=0$.  
When $0\le\theta \le\frac12\pi$, the spectrum is the positive continuum, 
and the generalised eigenfunctions are proportional to 
$\sin(k|x| +\delta_k)$ where $k>0$ and $\delta_k$ 
may be found in terms of $\theta$ from~\eqref{eq:app:robin}. 
When $\frac12\pi<\theta <\pi$, the spectrum consists of the 
positive continuum, with the generalised eigenfunctions as above, 
together with the single negative proper 
eigenvalue $- \cot^2(\theta)/L^2$ \cite{Bonneau:1999zq}. 

We may summarise: 
\begin{itemize}
\item 
On the odd subspace of $L_2(\BbbR \setminus\{0\})$, 
$- \Delta_\theta$ involves no boundary condition and 
coincides with the unique self-adjoint extension of 
$-\partial_x^2$ on the odd subspace of $L_2(\BbbR)$. 
\item 
On the even subspace of $L_2(\BbbR \setminus\{0\})$, 
$- \Delta_\theta$ is specified by the 
Robin boundary condition~\eqref{eq:app:robin}. 
\end{itemize}

The following two cases have special interest. 

When $\theta=\pi/2$, \eqref{eq:app:robin}
reduces to Neumann on each side of $x=0$. 
$- \Delta_{\pi/2}$ hence 
coincides with the essentially self-adjoint operator 
$- \partial_x^2$ on $L_2(\BbbR)$. There is no boundary condition and 
the point $x=0$ 
has no special role. 

When $\theta=0$, \eqref{eq:app:robin} 
reduces to Dirichlet on each side of $x=0$. Since the Dirichlet boundary condition is identically satisfied by odd wave functions, this means 
that $\BbbR_+$ and $\BbbR_-$ are 
completely decoupled by an impermeable two-sided 
Dirichlet wall at $x=0$. 

Finally, we note that when $\theta\ne0$, we may informally write
\begin{align}
- \Delta_\theta = -\partial_x^2 + {\frac{2 \cot\theta}{L}} \, \delta(x)
\ , 
\label{eq:dirac-delta-descr}
\end{align}
where $\delta(x)$ is Dirac's delta-function. 
In physics language, 
the boundary condition 
\eqref{eq:app:sa-gen} with \eqref{eq:app:U-theta}
can hence be described as a delta-function potential at $x=0$, 
with the $\theta$-dependent 
coefficient shown in~\eqref{eq:dirac-delta-descr}. 
Our reason to describe $- \Delta_\theta$ in terms of~$\theta$, 
rather than in terms of the coefficient 
of the Dirac delta in~\eqref{eq:dirac-delta-descr}, 
is that this will allow us to control in the main text 
the regularity of the Dirichlet limit $\theta\to0_+$, 
in which the coefficient of the Dirac delta in 
\eqref{eq:dirac-delta-descr} tends to~$+\infty$.

\section{Detector response in static half-space\label{app:statichalfspace}}

In this appendix we verify the properties quoted in 
subsection \ref{subsec:static-comparisons} about the response of 
the inertial detector \eqref{eq:detector-trajectory}
in Minkowski half-space with Dirichlet boundary conditions. 

In the Minkowski half-space $x>0$ with the Dirichlet boundary conditions at $x=0$, 
$\mathcal{W}(\tau',\tau'')$ consists of the 
Minkowski vacuum piece and the image contribution \cite{Juarez-Aubry:2014jba} 
\begin{align}
\Delta_{\text{Dir}}\mathcal{W}(\tau',\tau'') = 
\frac{1}{2\pi} \ln 
\Bigl[\mu \sqrt{(2d)^2 - (\tau'-\tau'' - i \epsilon)^2}  \, \Bigr] 
\ , 
\label{eq:deltaDirW1}
\end{align}
where $\epsilon \to 0_+$. 
From \eqref{eq:respfunc-altdef} and \eqref{eq:DirF1-decomp}
we then have 
\begin{align}
\Delta_{\text{Dir}} \mathcal{F}_1^{(1)}(\omega)
&= 
- \int_{-\infty}^\infty d\tau'' \, 
\overline{Q_\omega'(\tau'')}
\int_{-\infty}^\infty d\tau' 
Q_\omega(\tau')  \, \partial_{\tau'}
\Delta_{\text{Dir}}\mathcal{W}(\tau',\tau'')
\ . 
\label{eq:deltaDirF1-int1}
\end{align}
After inserting \eqref{eq:deltaDirW1} and writing out the $\tau'$-derivative, 
the inner integral may be evaluated using 
the identity 
$\lim_{\epsilon\to0_+}{(x-i\epsilon)}^{-1} = P(1/x) + i\pi \delta(x)$, 
where $P$ stands for the Cauchy principal value. 
Equations \eqref{eq:deltaDirF1-finalwithG}
in the main text then follow by writing out 
$\overline{Q_\omega'(\tau'')} = e^{i\omega\tau''} 
\bigl[\chi'(\tau'') + i\omega\chi(\tau'')\bigr]$ 
and performing straightforward integration variable changes. 

To obtain the large $|\omega|$ asymptotics, we assume $\omega\ne0$
and rewrite \eqref{eq:deltaDirF1-final} as 
\begin{align}
\Delta_{\text{Dir}} \mathcal{F}_1^{(1)}(\omega)
& = 
\frac{1}{2\pi} 
\Realpart \Biggl\{ e^{-2i\omega d}
\Biggl[
2 i \pi G_\omega(2d) \Theta(-\omega)
\notag
\\[1ex]
& \hspace{12ex}
+ 
\int_0^\infty ds 
\cos(\omega s) \, 
\frac{G_\omega(2d+s) - G_\omega(2d-s)}{s}
\notag
\\[1ex]
\
& \hspace{12ex}
-i 
\int_0^\infty ds 
\sin(\omega s) \, 
\frac{G_\omega(2d+s) + G_\omega(2d-s) - 2 G_\omega(2d)}{s}
\ \Biggr] \Biggr\}
\ , 
\label{eq:deltaDirF1-rearr}
\end{align}
adding and subtracting a term proportional to $G_\omega(2d)$ 
and using the identity 
$\int_0^\infty ds \, s^{-1} \sin(\omega s) 
= \tfrac12 \pi \sgn \omega$ where $\sgn$ is the signum function. 
The method of repeated integration by parts, 
integrating the trigonometric factor~\cite{wong}, 
shows that the integral terms in \eqref{eq:deltaDirF1-rearr} 
are $O^\infty\bigl(\omega^{-1}\bigr)$. 
Writing out $G_\omega(2d)$ gives formula 
\eqref{eq:DeltaDirF1-as} in the main text.

\section{Detector response for a rapidly created 
Dirichlet wall\label{app:lambdalimit}}

In this appendix we verify the properties quoted 
in subsection \ref{subsec:static-comparisons} 
about the response of the inertial detector \eqref{eq:detector-trajectory}
for a wall created in Minkowski 
space with the profile~\eqref{eq:infinite-profile}.

\subsection{Rapid wall creation limit}

At $x>0$, the spatially even mode functions are given by \eqref{eq:Umode-ansatz}
with~\eqref{eq:Esol-sampleprof}, while without the wall 
the spatially even mode functions are given by 
\eqref{eq:Umode-ansatz} with $E_k(t) = e^{-ikt}$. 
From \eqref{eq:F1lambda-decomp} we then have 
\begin{align}
\Delta \mathcal{F}_\lambda^{(1)}(\omega)
=
\int^{\infty}_{-\infty}\,d\tau'\,\int^{\infty}_{-\infty}\,d\tau''\, 
Q_\omega'(\tau') \overline{Q_\omega'(\tau'')} \, 
\Delta \mathcal{W}_\lambda (\tau',\tau'')
\ , 
\label{eq:DeltaFlambda-def}
\end{align}
where 
\begin{align}
\Delta \mathcal{W}_\lambda (\tau',\tau'') 
%&= \Delta \mathcal{W}_{\lambda,1} (\tau',\tau'') 
%+ 
%\Delta \mathcal{W}_{\lambda,2} (\tau',\tau'') 
%+ 
%\Delta \mathcal{W}_{\lambda,3} (\tau',\tau'') 
%\ , 
%\\[1ex]
%\Delta \mathcal{W}_{\lambda,1} (\tau',\tau'') 
&= 
\frac{1}{4\pi \bigl(1 + e^{-\lambda\tau'}\bigr) \bigl(1 + e^{-\lambda\tau''}\bigr)}
\Bigl[ 2 E_1\bigl(\mu(\epsilon + i\Delta\tau)\bigr)
\notag
\\
& 
\hspace{12ex}
- e^{-\lambda\Delta\tau} E_1\bigl((\mu+i\lambda)(\epsilon + i\Delta\tau)\bigr)
- e^{\lambda\Delta\tau} E_1\bigl((\mu-i\lambda)(\epsilon + i\Delta\tau)\bigr)
\Bigr]
\notag
\\[1ex]
& \hspace{2ex}
- \frac{1}{4\pi \bigl(1 + e^{-\lambda\tau'}\bigr)}
\Bigl[ 
E_1\bigl(\mu(\epsilon + i\Delta\tau)\bigr)
+ E_1\bigl(\mu(\epsilon + i(\Delta\tau -2d))\bigr)
\notag
\\
& 
\hspace{20ex}
- e^{-\lambda\Delta\tau} E_1\bigl((\mu+i\lambda)(\epsilon + i\Delta\tau)\bigr)
\notag
\\
& 
\hspace{20ex}
- e^{-\lambda(\Delta\tau -2d)} E_1\bigl((\mu+i\lambda)(\epsilon + i(\Delta\tau -2d))\bigr)
\Bigr]
\notag
\\[1ex]
& \hspace{2ex}
- \frac{1}{4\pi \bigl(1 + e^{-\lambda\tau''}\bigr)}
\Bigl[ 
E_1\bigl(\mu(\epsilon + i\Delta\tau)\bigr)
+ E_1\bigl(\mu(\epsilon + i(\Delta\tau +2d))\bigr)
\notag
\\
& 
\hspace{20ex}
- e^{\lambda\Delta\tau} E_1\bigl((\mu-i\lambda)(\epsilon + i\Delta\tau)\bigr)
\notag
\\
& 
\hspace{20ex}
- e^{\lambda(\Delta\tau +2d)} E_1\bigl((\mu-i\lambda)(\epsilon + i(\Delta\tau +2d))\bigr)
\Bigr]
\ , 
\label{eq:DeltaWlambda-E1}
\end{align}
$\Delta\tau := \tau'-\tau''$, the positive constant $\mu$ is an infrared cutoff, and 
$E_1$ is the exponential integral in the notation of~\cite{dlmf}, 
taking values on its principal branch in the sense of $\epsilon\to0_+$. 

We wish to take the limit $\lambda\to\infty$ in~\eqref{eq:DeltaFlambda-def}. 
For the terms in 
\eqref{eq:DeltaWlambda-E1} that contain $\lambda$ in the argument of~$E_1$, 
we may use properties of $E_1$ from \cite{dlmf}
[the integral representation
(6.2.1) and the asymptotic expansion (6.12.1)]
to show that the contribution from these terms vanishes in the limit 
$\lambda\to\infty$. For the remaining terms in 
\eqref{eq:DeltaWlambda-E1} the limit is elementary, leading to equations 
\eqref{eq:DeltaF-def} and \eqref{eq:DeltaW-def} in the main text.

\subsection{Simplified expression \eqref{eq:deltaF1-finalwithH} 
for the response function}

We now express $\Delta \mathcal{F}^{(1)}$, given by 
\eqref{eq:DeltaF-def} with~\eqref{eq:DeltaW-def}, in terms of integrals 
that do not involve special functions. 

Starting from \eqref{eq:DeltaF-def} with \eqref{eq:DeltaW-def} 
and breaking the integrations into subdomains gives 
\begin{subequations}
\begin{align}
\Delta \mathcal{F}^{(1)}(\omega)
&=
\Delta \mathcal{F}_1^{(1)}(\omega) + \Delta \mathcal{F}_2^{(1)}(\omega)
\ , 
\\[1ex]
\Delta \mathcal{F}_1^{(1)}(\omega)
&= 
- \frac{1}{2\pi}
\Realpart 
\int_0^\infty d\tau' \int_{-\infty}^0 d\tau'' \, 
Q_\omega'(\tau') \overline{Q_\omega'(\tau'')} \, E_1 \bigl(i \mu (\tau'-\tau'')\bigr)
\ , 
\label{eq:DeltaF1-raw}
\\[1ex]
\Delta \mathcal{F}_2^{(1)}(\omega)
&= 
- \frac{1}{2\pi}
\Realpart 
\int_0^\infty d\tau' \int_{-\infty}^\infty d\tau'' \, 
Q_\omega'(\tau') \overline{Q_\omega'(\tau'')} \, E_1 \bigl(i \mu (\tau'-\tau'' - 2d)\bigr)
\ , 
\label{eq:DeltaF2-raw}
\end{align}
\end{subequations}
where $E_1$ takes values on its principal branch. 

Consider $\Delta \mathcal{F}_1^{(1)}$. In \eqref{eq:DeltaF1-raw}, 
interchanging the integrals and integrating by parts in the inner integral gives 
\begin{subequations}
\begin{align}
\Delta \mathcal{F}_1^{(1)}(\omega)
&= 
\Delta \mathcal{F}_{1,1}^{(1)}(\omega)
+ \Delta \mathcal{F}_{1,2}^{(1)}(\omega)
\ , 
\\
\Delta \mathcal{F}_{1,1}^{(1)}(\omega)
&= 
\frac{\chi(0)}{2\pi}
\Realpart 
\int_{-\infty}^0 d\tau'' \, 
\overline{Q_\omega'(\tau'')} \, E_1 (-i \mu \tau'')
\ , 
\label{eq:DeltaF1-1}
\\
\Delta \mathcal{F}_{1,2}^{(1)}(\omega)
& = 
- \frac{1}{2\pi}
\Realpart 
\int_{-\infty}^0 d\tau'' \, \overline{Q_\omega'(\tau'')}
\int_0^\infty d\tau'  \, 
Q_\omega(\tau')  \, \frac{e^{-i\mu(\tau'-\tau'')}}{(\tau'-\tau'')}
\label{eq:DeltaF1-2}
\ . 
\end{align}
\end{subequations}

From now on we assume $\omega+\mu\ne0$. 

To evaluate $\Delta \mathcal{F}_{1,1}^{(1)}$, 
we write $\Delta \mathcal{F}_{1,1}^{(1)} = \lim_{\epsilon\to0_+} \Delta \mathcal{F}_{1,1,\epsilon}^{(1)}$, where 
$\Delta \mathcal{F}_{1,1,\epsilon}^{(1)}$ is as 
\eqref{eq:DeltaF1-1} but with the upper limit of integration replaced by~$-\epsilon$. 
Integrating by parts, renaming the integration variable by $\tau'' = -s$, and adding and subtracting under the integral a term proportional to $e^{-i(\omega+\mu)s}/s$, we find 
\begin{align}
\Delta \mathcal{F}_{1,1,\epsilon}^{(1)}(\omega)
&= 
\frac{\chi(0)}{2\pi} 
\Biggl\{
\int_\epsilon^\infty ds \, 
\frac{\chi(0) - \chi(-s)}{s} 
\cos\bigl((\omega+\mu)s\bigr)
\notag
\\
& \hspace{3ex}
+ \Realpart \left[ \chi(-\epsilon) e^{-i\omega\epsilon} E_1(i\mu\epsilon)\right]
\ - \chi(0)\int_\epsilon^\infty 
\frac{ds}{s} \cos\bigl((\omega+\mu)s\bigr)
\Biggr\}
\ . 
\label{eq:DeltaF1-1-eps}
\end{align}
The last term in \eqref{eq:DeltaF1-1-eps} is proportional to 
the cosine integral $\Ci\bigl(|\omega+\mu|\epsilon\bigr)$~\cite{dlmf}, 
and the limit $\epsilon \to0_+$ can be taken using the small 
argument asymptotic forms of $\Ci$ and $E_1$~\cite{dlmf}. We find 
\begin{align}
\Delta \mathcal{F}^{(1)}_{1,1}(\omega)
& = 
\frac{{[\chi(0)]}^2}{2\pi}
\ln\bigl|1 + (\omega/\mu)\bigr|
\ + 
\frac{\chi(0)}{2\pi} 
\int_0^\infty ds \, 
\frac{\chi(0) - \chi(-s)}{s} 
\cos\bigl((\omega+\mu)s\bigr)
\ . 
\label{eq:DeltaF1-1-final}
\end{align}

To evaluate $\Delta \mathcal{F}_{1,2}^{(1)}$, 
we interchange the integrals in~\eqref{eq:DeltaF1-2}, write 
$\overline{Q_\omega'(\tau'')} = e^{i\omega\tau''} 
\bigl[\chi'(\tau'') + i\omega\chi(\tau'')\bigr]$, 
change the integration variable in the inner integral by $\tau'' = \tau'-s$ 
where $s \in (\tau',\infty)$, and interchange the integrals again. 
Renaming $\tau'$ as $u$, we find 
\begin{align}
\Delta \mathcal{F}^{(1)}_{1,2}(\omega)
& = 
- 
\frac{1}{2\pi} 
\int_0^\infty ds 
\cos\bigl((\omega+\mu)s\bigr)
\, s^{-1} \int_0^s du \, \chi(u) \chi'(u-s)
\notag
\\[1ex]
& 
\hspace{3ex}
- 
\frac{\omega}{2\pi} 
\int_0^\infty ds 
\sin\bigl((\omega+\mu)s\bigr)
\, s^{-1} \int_0^s du \, \chi(u) \chi(u-s)
\ . 
\label{eq:DeltaF1-2-final}
\end{align}

Consider next $\Delta \mathcal{F}_2^{(1)}$. We 
change the integration variable in the inner integral 
in \eqref{eq:DeltaF2-raw}
by $\tau'' = \tau' - 2d -s$, 
interchange the integrals and rename $\tau'$ as~$u$, 
obtaining 
\begin{align}
\Delta \mathcal{F}_2^{(1)}(\omega)
& = 
- \frac{1}{2\pi}
\Realpart 
\int_{-\infty}^\infty ds \, E_1 (i \mu s)
\int_0^\infty du \, Q_\omega'(u) \overline{Q_\omega'(u - 2d -s)}
\notag
\\[1ex]
& = 
\frac{1}{2\pi}
\Realpart \left\{ e^{-2i\omega d}
\int_{-\infty}^\infty ds \, E_1 (i \mu s)
\, \frac{d}{ds} \bigl[ e^{-i\omega s}H_\omega(2d+s) \bigr]
\right\}
\ , 
\label{eq:DeltaF2-mat1}
\end{align}
where $H_\omega$ is given by~\eqref{eq:H-def}. 
The last equality in 
\eqref{eq:DeltaF2-mat1} follows by observing that 
$\int_0^\infty du \, Q_\omega'(u) \overline{Q_\omega'(u - 2d -s)}
= - \frac{d}{ds} \int_0^\infty du \, Q_\omega'(u) \overline{Q_\omega(u - 2d-s)}$. 

We may now write 
$\Delta \mathcal{F}_{2}^{(1)} = \lim_{\epsilon\to0_+}\Delta \mathcal{F}_{2,\epsilon}^{(1)}$, 
where $\Delta \mathcal{F}_{2,\epsilon}^{(1)}$
is as in \eqref{eq:DeltaF2-mat1} except that the integration over 
$s$ omits the interval 
$(-\epsilon, \epsilon)$. Integration by parts gives 
\begin{align}
\Delta \mathcal{F}_{2,\epsilon}^{(1)}(\omega)
& = 
\frac{1}{2\pi} 
\Realpart \Biggl\{ e^{-2i\omega d}
\Biggl[
E_1(-i\mu\epsilon) e^{i\omega\epsilon} H_\omega(2d-\epsilon)
- E_1(i\mu\epsilon) e^{-i\omega\epsilon} H_\omega(2d+\epsilon)
\notag
\\[1ex]
& \hspace{10ex}
+ 
\int_{-\infty}^{-\epsilon} ds \, 
\frac{e^{-i(\mu+\omega) s}}{s} H_\omega(2d+s) 
+ 
\int_\epsilon^\infty ds \, 
\frac{e^{-i(\mu+\omega) s}}{s} H_\omega(2d+s) 
\ \Biggr] \Biggr\}
\ . 
\label{eq:temp3}
\end{align}
The limit $\epsilon\to0$ 
in \eqref{eq:temp3} can be taken by using 
(6.2.4) of \cite{dlmf} in the first two terms 
and by the change of variables $s \to -s$ in the third term. We find 
\begin{align}
\Delta \mathcal{F}_2^{(1)}(\omega)
& = 
\frac{1}{2\pi} 
\Realpart \Biggl\{ e^{-2i\omega d}
\Biggl[
i \pi H_\omega(2d)
\notag
\\[1ex]
& \hspace{11ex}
+ 
\int_0^\infty ds \, 
\frac{e^{-i(\mu+\omega) s} H_\omega(2d+s) - e^{i(\mu+\omega) s} H_\omega(2d-s)}{s}
\ \Biggr] \Biggr\}
\ . 
\label{eq:deltaF-2-app-final}
\end{align}

Combining 
\eqref{eq:DeltaF1-1-final}, 
\eqref{eq:DeltaF1-2-final}
and~\eqref{eq:deltaF-2-app-final}, we obtain 
formula \eqref{eq:deltaF1-finalwithH} in the main text.

\subsection{Limit of large energy gap}

We now obtain the large $|\omega|$ form of $\Delta \mathcal{F}^{(1)}(\omega)$. 

For $\Delta \mathcal{F}_1^{(1)}$, we apply to the 
integral terms in \eqref{eq:DeltaF1-1-final} and 
\eqref{eq:DeltaF1-2-final} the method of repeated integration by parts, 
integrating the trigonometric factor~\cite{wong}. 
The second integral term in \eqref{eq:DeltaF1-2-final} is 
$- \frac{{[\chi(0)]}^2}{2\pi} + O\bigl(\omega^{-1}\bigr)$ 
and all the other integral terms are $O\bigl(\omega^{-1}\bigr)$. 
Combining, we have 
\begin{align}
\Delta \mathcal{F}_1^{(1)}(\omega)
& = 
\frac{{[\chi(0)]}^2}{2\pi}
\ln\bigl( e^{-1} \bigl|1 + (\omega/\mu)\bigr| \bigr) 
+ O\bigl(\omega^{-1}\bigr)
\ . 
\end{align}

For $\Delta \mathcal{F}_2^{(1)}$, we apply to \eqref{eq:deltaF-2-app-final} 
the same method that was applied to \eqref{eq:deltaDirF1-final}
in Appendix~\ref{app:statichalfspace}. We obtain 
\begin{align}
\Delta \mathcal{F}_2^{(1)}(\omega)
& = \Theta(-\omega)
\Realpart \bigl[ i e^{-2i\omega d} H_\omega(2d) \bigr] 
\ + O^\infty\bigl(\omega^{-1}\bigr)
\ .  
\label{eq:deltaF-2-app-rearr}
\end{align}

Combining these observations and writing out $H_\omega(2d)$ gives formula 
\eqref{eq:DeltaF1-as} in the main text.

\section{Bogoliubov coefficients\label{app:bogocoeffs}}

In this appendix we examine briefly the wall creation in terms of 
Bogoliubov coefficients. 
We anticipate that this formalism will be useful  
for analysing the entanglement structure in the bursts 
of particles that the wall 
formation generates~\cite{Brown:2014qna}. 

\subsection{Wall in Minkowski space}

Consider the wall creation in Minkowski spacetime, 
in the notation of Section~\ref{sec:Minkowski}. 
We recall that it suffices to consider the spatially even mode functions, 
and we write down the formulas for the mode functions 
only in the half-space $x>0$. 

The mode functions that reduce to standard Minkowski 
mode functions before the wall starts to form are denoted by 
$U_k$ with $k>0$ and are given by 
\eqref{eq:Umode-ansatz}
with \eqref{eq:E-intermsof-R}--\eqref{eq:Bscaled-diffeq}. 
The mode functions that reduce to standard 
Minkowski mode 
functions with the Dirichlet boundary condition 
after the wall has fully formed are denoted by $W_k$ with $k>0$ 
and are given by 
\begin{align}
W_k(u,v) = \frac{1}{\sqrt{8\pi k}} 
\left[
\tilde E_k (v) - e^{-iku}
\right]
\ , 
\label{eq:Wmode-ansatz}
\end{align}
where 
\begin{align}
\tilde E_k (v) = e^{-ikv}
\hspace{3ex}
\text{for $v \ge \lambda^{-1}$}
\ , 
\label{eq:Etilde-latesol}
\end{align}
and the expression for $\tilde E_k (v)$ for $v < \lambda^{-1}$ 
can be found by the methods of Section~\ref{sec:Minkowski} 
but will not be needed here. 
$W_k$~satisfy the Klein-Gordon orthonormality 
relations similar to~\eqref{eq:U-KG-orthonormality}. 

We write the Bogoliubov transformation between the two 
sets of modes in the notation of 
\cite{birrell-davies} as 
\begin{align}
W_{k} = \int_0^{\infty} \left(\alpha_{k l} U_{l} 
+ \beta_{k l} \overline{U_{l}} \, \right) 
\, dl 
\ , 
\end{align}
so that 
\begin{align}
\alpha_{kl} = \bigl(W_k, U_{l}\bigr)
\ , \ \ \ \ 
\beta_{kl} = - \bigl(W_k, \overline{U_{l}} \, \bigr)
\ . 
\label{eq:Bogo-alpha-beta}
\end{align}
The inner products in \eqref{eq:Bogo-alpha-beta} may be evaluated by 
choosing a hypersurface of constant $t$ at $t \ge \lambda^{-1}$ and
using 
\eqref{eq:Umode-ansatz}
with \eqref{eq:E-intermsof-R}--\eqref{eq:R-sol-full}
and \eqref{eq:Wmode-ansatz} with~\eqref{eq:Etilde-latesol}. 
The result is 
\begin{subequations}
\label{eq:Bogo-sol-alphabeta}
\begin{align}
\alpha_{kl} &= 
\frac{1}{2\pi}
\sqrt{\frac{k}{l}}
\left[
i \, P \left( \frac{1 + e^{i(l-k)/\lambda}}
{l-k}\right)
- \frac{1}{\lambda}
\int_0^1 e^{-i(k/\lambda)y} \, \overline{R_{l/\lambda}(y)} \, dy 
\right]
\ , 
\label{eq:Bogo-sol-alpha}
\\[1ex]
\beta_{kl} &= 
\frac{1}{2\pi}
\sqrt{\frac{k}{l}}
\left[
i \, \frac{1 + e^{-i(l+k)/\lambda}}
{l+k}
+ \frac{1}{\lambda}
\int_0^1 e^{-i(k/\lambda)y} \, R_{l/\lambda}(y) \, dy 
\right]
\ , 
\label{eq:Bogo-sol-beta}
\end{align}
\end{subequations}
were $P$ denotes the Cauchy principal value. 
The presence of particle creation is manifest in the nonvanishing 
beta-coefficients~\eqref{eq:Bogo-sol-beta}. 

\subsection{Wall in the Dirichlet cavity}

For the wall creation in the Dirichlet cavity we may proceed similarly, 
in the notation of Section~\ref{sec:cavity}. 
We assume the cavity to be so large that $a> 2/\lambda$. 

For $\lambda^{-1} < t < a/2$, 
combining the results of Sections \ref{sec:Minkowski} and 
\ref{sec:cavity} shows that the $V$-modes \eqref{eq:Vmode-ansatz} 
are obtained from the $U$-modes of 
\eqref{eq:Umode-ansatz}
with \eqref{eq:E-intermsof-R}--\eqref{eq:Bscaled-diffeq} 
by including the overall multiplicative factor 
$\sqrt{2\pi/a}$ and restricting $ka/\pi$ to odd positive integers, 
while the $W$-modes are obtained from 
\eqref{eq:Wmode-ansatz} with 
\eqref{eq:Etilde-latesol}
by including the overall multiplicative factor 
$\sqrt{2\pi/a}$ and restricting $ka/\pi$ to even positive integers. 
The Bogoliubov transformation is written as in 
\eqref{eq:Bogo-alpha-beta} but the integral replaced by a sum. 
It follows that the Bogoliubov coefficients are obtained from 
\eqref{eq:Bogo-sol-alphabeta} by including 
the overall multiplicative factor 
$2\pi/a$, restricting $ka/\pi$ to even positive integers, 
restricting $la/\pi$ to odd positive integers, and dropping the symbol~$P$. 
Again, the presence of particle creation is manifest in the nonvanishing 
beta-coefficients.

\end{appendix}

\end{document}